\documentclass[twocolumn, showkeys, aps, pra, showpacs, superscriptaddress, floatfix, nofootinbib, 10pt]{revtex4-2}

\usepackage{amsmath,bbm,amsthm}
\usepackage{amssymb}
\usepackage{graphicx}
\usepackage{dsfont}
\usepackage{cancel}
\usepackage{enumitem}   
\usepackage{comment}
\usepackage{tabularx}
\usepackage{booktabs}
\usepackage{array}
\usepackage{multirow}
\usepackage{ragged2e}
\usepackage{tikz}

\usepackage{hyperref}
\hypersetup{colorlinks,allcolors=blue}

\makeatletter
\newtheorem*{rep@theorem}{\rep@title}
\newcommand{\newreptheorem}[2]{%
\newenvironment{rep#1}[1]{%
 \def\rep@title{#2 \ref{##1}}%
 \begin{rep@theorem}}%
 {\end{rep@theorem}}}
\makeatother

\newreptheorem{thm}{Theorem}

\newtheorem{result}{Result}

\renewcommand{\today}{\number\day\space\ifcase\month\or
   January\or February\or March\or April\or May\or June\or
   July\or August\or September\or October\or November\or December\fi
   \space\number\year}

\begin{document}

\title{Quantum-informed learning of genuine network nonlocality beyond idealized resources}

\author{Anantha Krishnan Sunilkumar}
\email{readatanantha@gmail.com}
\affiliation{School of Physics, IISER Thiruvananthapuram, Kerala 695551, India}
\author{Anil Shaji}
\affiliation{School of Physics, IISER Thiruvananthapuram, Kerala 695551, India}
\affiliation{Center for High Performance Computing, IISER Thiruvananthapuram, Kerala 695551, India}
\author{Debashis Saha}
\affiliation{School of Physics, IISER Thiruvananthapuram, Kerala 695551, India}
\affiliation{Department of Physics, School of Basic Sciences, IIT Bhubaneswar, Odisha 752050, India}

\date{\today}

\begin{abstract}

We address the characterization of genuine network nonlocal correlations, which remain highly challenging due to the non-convex nature of local correlations even in the distinct triangle scenario with three sources and three observers implementing one four-outcome measurement. We introduce a scalable causally inferred Bayesian learning framework called the \textit{Layered Local Hidden Variable Neural Network (Layered LHV-Net)} to learn the local statistics in network Bell tests. Using this framework, we identify a new class of measurement settings that exhibit the most robust nonlocality compared to previously known measurements. Remarkably, our study shows that the nonlocality measure becomes non-zero only when the visibility of the shared Bell state exceeds $0.94$, surpassing previously reported noise robustness thresholds. Further, we examine correlations where shared states originate from dissimilar sources, finding that nonlocality is observed only if all the involved states are sufficiently entangled. Finally, we analyze a scenario in which the sources are allowed to share classical randomness. We find that nonlocal correlations persist even when the sources share up to $3$ units of randomness, whereas a local model reproducing the quantum correlations only becomes possible when $4$ units of shared randomness are available. Apart from the results, the work succeeds in showing that quantum-informed machine learning approaches as foundational frameworks can greatly benefit the field of quantum information.

\end{abstract}

\maketitle

\section{Introduction}

The study of quantum correlations unveils profound aspects of nature, bridging the foundations of quantum theory with applications in quantum technologies. From questions raised on quantum theory by Einstein, Podolsky, and Rosen (EPR) \cite{Can_Quantum-Mechanical_Description_of_Physical_Reality_Be_Considered_Complete?EPR} through Bell's theorem \cite{On_the_Einstein_Podolsky_Rosen_paradox:bell_einstein_1964}  and its applications to real-world technologies \cite{Ekert:PhysRevLett.67.661}, massive progress has been realized over the years. Bell’s seminal theorem presented nonlocality by demonstrating that entangled states can exhibit correlations beyond the constraints placed by any local realistic theory. Experimental realizations of Bell nonlocality \cite{Bellnonlocality:RevModPhys.86.419}, for instance, in quantum steering highlight the inherent quantum effects in entangled states, which serve as a pivotal resource for device-independent quantum technologies \citep{No_signalling_and_Quantum_Key_Distribution:PhysRevLett.95.010503, Device-Independent_Security_of_Quantum_Cryptography_against_Collective_Attacks:PhysRevLett.98.230501, colbeck2011quantumrelativisticprotocolssecure, Random_numbers_certified_by_Bell’s_theorem:Pironio_2010}.

The Clauser-Horne-Shimony-Holt (CHSH) noise-robust proof \cite{Proposed_CHSH_experiment:PhysRevLett.23.880} is a cornerstone on which many of these achievements have been built, paving the way for the experiments \citep{Loophole-free_Bell_inequality_violation_using_electron_spins_separated_by_1.3_kilometres:Hensen2015, Strong_Loophole-Free_Test_of_Local_Realism:PhysRevLett.115.250402, Significant-Loophole-Free_Test_of_Bell's_Theorem_with_Entangled_Photons:PhysRevLett.115.250401} empirically confirming quantum violation and also giving rise to the device-independent (DI) paradigm \citep{No_signalling_and_Quantum_Key_Distribution:PhysRevLett.95.010503, Device-Independent_Security_of_Quantum_Cryptography_against_Collective_Attacks:PhysRevLett.98.230501, colbeck2011quantumrelativisticprotocolssecure, Random_numbers_certified_by_Bell’s_theorem:Pironio_2010}.

\begin{figure}[t]
\centering

\begin{tikzpicture}
\node[anchor=south west, inner sep=0] (img) 
  {\includegraphics[width=1.0\columnwidth]{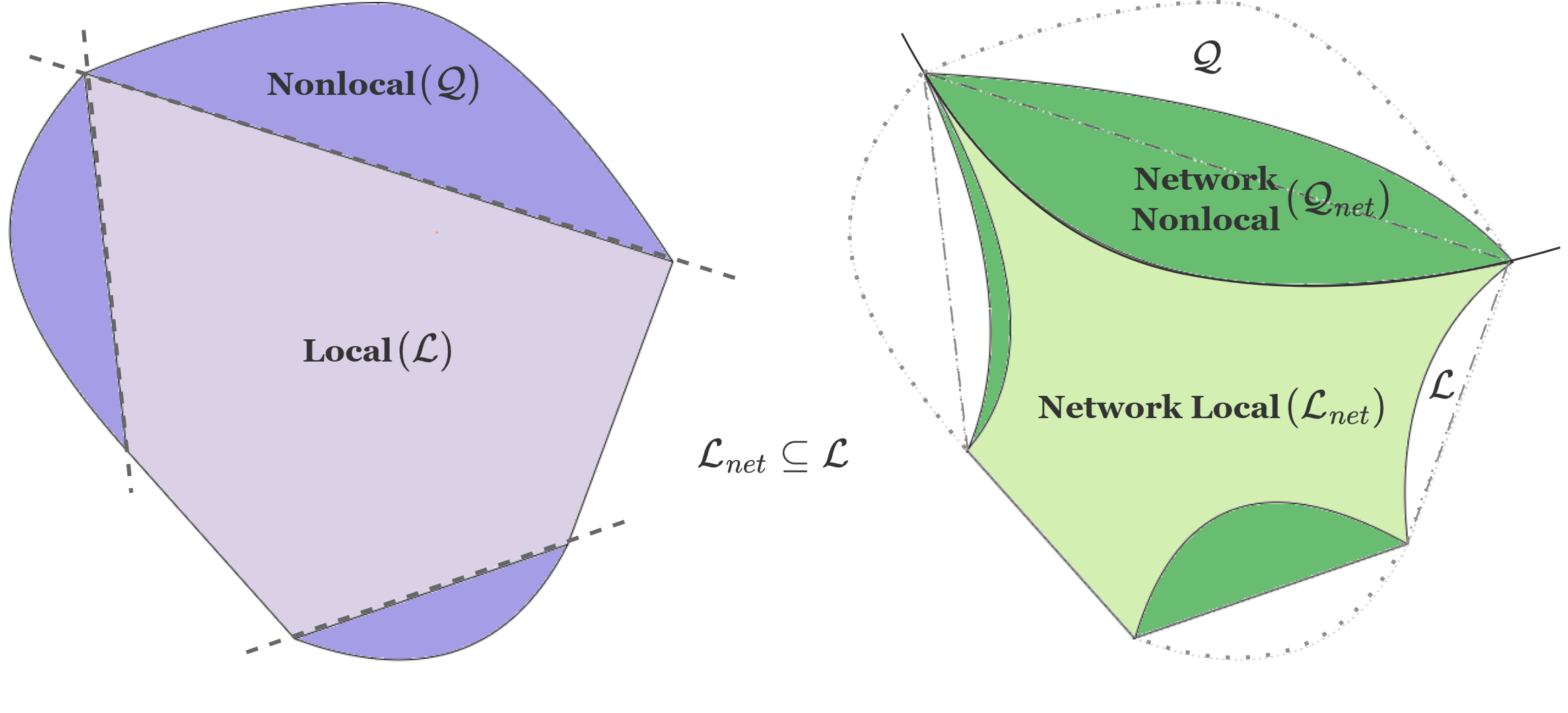}};
\begin{scope}[x={(img.south east)},y={(img.north west)}]
  \node at (0.05,0.02) {\scriptsize{\textbf{(a)}}};
  \node at (0.55,0.02) {\scriptsize{\textbf{(b)}}};
\end{scope}

\end{tikzpicture}
\vspace{-0.7cm}
\caption{\justifying The set of correlations. (a) The set of standard local correlations is contained in the set of standard non-local correlations without independent sources $(\mathcal{L} \subsetneq \mathcal{Q})$ \cite{Bellnonlocality:RevModPhys.86.419} witnessed by violating the standard Bell inequalities ($\rule[.5ex]{1em}{.4pt}\,\rule[.5ex]{.2em}{.4pt}\,\rule[.5ex]{1em}{.4pt}$). (b) The set of network local correlations is contained in the set of standard local correlations $(\mathcal{L_\text{net}} \subseteq \mathcal{L})$ as well as the set of network non-local correlations $(\mathcal{L_\text{net}} \subsetneq \mathcal{Q_\text{net}})$ (witnessed by violating the LHV network descriptions~(\ref{eq:network_trilocal_model}) ($\rule[.5ex]{2em}{.4pt}$)) which in turn is contained in the set of standard non-local correlations $(\mathcal{Q_\text{net}} \subsetneq \mathcal{Q})$.}
\label{fig:Local_and_nonlocal_domain}
 \vspace{-0.25cm}
\end{figure}

Recently, research into decentralized multipartite networks \citep{Characterizing_the_Nonlocal_Correlations_Created_via_Entanglement_Swapping:PhysRevLett.104.170401, Bilocal_versus_nonbilocal_correlations_in_entanglement-swapping_experiments:PhysRevA.85.032119, Beyond_Bell’s_theorem:_correlation_scenarios:Fritz_2012, Quantum_Correlations_Take_a_New_Shape:Pusey2019} without inputs uncovered new layers of complexity (see Fig.~\ref{fig:Local_and_nonlocal_domain}),  introducing entangled measurements as a resource alongside entangled states as in standard Bell scenarios \cite{Bell_nonlocality_in_networks:Tavakoli2022}. The causal scenarios unfolding in the network owing to the entangling measurements bring new nuances to the standard Bell scenario. In this line of studies, there exist distributions from network scenarios that can be viewed as a clever embedding of a standard Bell test derived by Fritz \cite{Beyond_Bell’s_theorem:_correlation_scenarios:Fritz_2012}, and as such, only rely on entangled states. For this class of distributions, noise-robust proofs \cite{Causal_Networks_and_Freedom_of_Choice_in_Bell's_Theorem:PRXQuantum.2.040323} have been developed, leading to their first experimental tests \cite{Experimental_nonclassicality_in_a_causal_network_without-assuming_freedom_of_choice:Polino_2023} (see also \cite{rohrlich1995nonlocalityaxiomquantumtheory}). Now, unlike these, there exist constructions that bring novel classes of quantum correlations that cannot be traced back to standard Bell scenarios and are unique to network scenarios, demonstrating a genuine network nonlocality \citep{Bilocal_versus_nonbilocal_correlations_in_entanglement-swapping_experiments:PhysRevA.85.032119, Renou_Genuine2019:PhysRevLett.123.140401}. Here, entangled measurements play a key role in conjunction with entangled states, as shown by self-testing-focused approaches \citep{mayers2004selftestingquantumapparatus, Self-Testing_Entangled_Measurements_in_Quantum_Networks:Renou_2018, Noise-Resistant_Device-Independent_Certification_of_Bell_State_Measurements:PhysRevLett.121.250506, Genuine_network_quantum_nonlocality_and_self-testing:_upi__2022, sekatski2022partialselftestingrandomnesscertification}. While progress has been made in studying these correlations, all current conclusive proofs are noiseless \citep{Renou_Genuine2019:PhysRevLett.123.140401, Towards_a_minimal_example_of_quantum_nonlocality_without_inputs:Boreiri_2023, Network_nonlocality_via_rigidity_of_token_counting_and_color_matching:Renou_2022, Single-photon_nonlocality_in_quantum_networks:Abiuso_2022, Proofs_of_Network_Quantum_nonlocality_in_Continuous_Families_of_Distributions:Pozas_Kerstjens_2023}. That is, they consider a setting where the shared states are pure with joint entangled measurements. Hence, addressing noise robustness and genuine network correlations within mixed states conclusively is an important open problem which we address in this work. This will be a significant step towards experimental realization and possible applications of such correlations.

Unlike the standard Bell scenario, multipartite networks of our interest that exhibit genuine nonlocality have non-convex local boundaries owing to their source independence, making optimization a hard problem. Due to the proximity of these distributions to the local landscape, existing studies \citep{Theory-independent_limits_on_correlations_from_generalized_Bayesian_networks:Henson2014, Nonlocal_correlations_in_the_star-network_configuration:PhysRevA.90.062109, Causal_structures_from_entropic_information:_geometry_and_novel_scenarios:Chaves2014, chaves2014inferringlatentstructuresinformation, Information–theoretic_implications_of_quantum_causal_structures:Chaves_2015, Nonlinear_Bell_Inequalities_Tailored_for_Quantum_Networks:PhysRevLett.116.010403, Polynomial_Bell_Inequalities:PhysRevLett.116.010402, The_Inflation_Technique_Completely_Solves_the_Causal_Compatibility_Problem:Navascu_s_2020, Universal_bound_on_the_cardinality_of_local_hidden_variables_in_networks:Rosset_2017, Causal_compatibility_inequalities_admitting_quantum_violations_in_the_triangle_structure:PhysRevA.98.022113, Non-Shannon_inequalities_in_the_entropy_vector_approach_to_causal_structures:Weilenmann2018nonshannon, Computationally_Efficient_Nonlinear_Bell_Inequalities_for_Quantum_Networks:PhysRevLett.120.140402, The_Inflation_Technique_for_Causal_Inference_with_Latent_Variables:Wolfe_2019, Renou_Genuine2019:PhysRevLett.123.140401, Constraints_on_nonlocality_in_networks_from_no-signaling_and_independence:Gisin2020, Limits_on_Correlations_in_Networks_for_Quantum_and_No-Signaling_Resources:PhysRevLett.123.070403, Bounding_the_Sets_of_Classical_and_Quantum_Correlations_in_Networks:PhysRevLett.123.140503, Semidefinite_programming_relaxations_for_quantum_correlations:Tavakoli_2024, Quantum_Inflation:A_General_Approach_to_Quantum_Causal_Compatibility:Wolfe_2021, Genuine_Multipartite_nonlocality_Is_Intrinsic_to_Quantum_Networks:Contreras_Tejada_2021} have mainly explored ideal scenarios with symmetric distributions, such as the Elegant scenario \cite{Entanglement_25_years_after_quantum_teleportation:Gisin2019,Bilocal_Bell_Inequalities_Violated_by_the_Quantum_Elegant_Joint_Measurement:PhysRevLett.126.220401}, also paving the way for experiments in such limited regimes \cite{Demonstrating_the_power_of_quantum_computers_certification_of_highly_entangled_measurements_and_scalable_quantum_nonlocality:Baumer_2021, wang2024experimentalgenuinequantumnonlocality}. While methods like self-testing \cite{Genuine_network_quantum_nonlocality_and_self-testing:_upi__2022} and other frameworks, including inflation and causal inference strategies  \cite{The_Inflation_Technique_for_Causal_Inference_with_Latent_Variables:Wolfe_2019} explore nonlocality and network scenarios rigorously, they are either limited to ideal scenarios, are computationally expensive with increasing complexity, or fail to accurately map the boundary of these correlations, leaving the nature of these correlations in terms of their network topologies, resources, and noise robustness ambiguous.

Recent advances in interdisciplinary fields with machine learning, quantum foundations, and causal inference \cite{Machine_Learning_Nonlocal_Correlations:PhysRevLett.122.200401, Machine_Learning_Detection_of_Bell_nonlocality_in_Quantum_Many-Body-Systems:PhysRevLett.120.240402, Tamas_2020, Probabilistic_Graphical_Models:Koller1989, Learning_Functional_Causal_Models_with_Generative_Neural_Networks:Goudet2018, Causality:Hitchcock2001} provide a strong basis for addressing these limitations of traditional numerical and analytical techniques. Inspired by these approaches, we introduce an operational causally inferred Bayesian learning framework called the Layered LHV-Net framework, where we introduce the rank of the quantum state as a previously untapped degree of freedom in learning the local statistics of Bell tests. This new framework successfully addresses the inconsistencies with earlier methods, providing a noise-robust proof for quantum multipartite networks. Applying this framework to the triangle scenario, we uncover interesting details on genuine network nonlocal correlations. We confirm the consistency of our framework with existing results and benchmark our method based on new, elegant limits to robustness, achieving higher accuracy with fewer resources. 

Using the framework, we study genuine network nonlocal correlations of a more general form of the shared state between the parties and a general form of the measurements that was not considered before. Our results reveal a new set of non-maximal measurements giving the best value of nonlocality. We also find that the nonlocal correlations persist only close to pure state distributions, with a new limit of $0.94$ on noise robustness based on Werner noise, emphasizing that these correlations are much more limited in scope and stricter than previously understood. Additionally, we find that these correlations require all sources to send states of a certain degree of entanglement for maximal nonlocality; alongside this, we find a separate class of non-maximally entangled states distinct from Werner states capable of exhibiting these correlations.

Further, by adapting our framework to introduce shared randomness, we find that genuine network nonlocal correlations exhibit robustness to shared randomness even with noise. We also provide the numerical response functions of the LHV descriptions for the $v = 0.94$ robustness parameter, previously considered genuine network nonlocal.

Our method can be generalized to other causal structures, providing a definitive noise-robust proof and a framework to study Genuine Network Nonlocal (GNN) correlations. Our contribution not only gives novel results but also shows that machine learning algorithms, unlike those typically associated with data-driven prediction, when structurally embedded with domain-specific knowledge and principles, can greatly contribute to quantum foundational research, elevating it from a predictive tool to a foundational framework.

\section{Genuine network nonlocality in the triangle scenario}

\vspace{-0.4cm}
\begin{figure}[ht]
    \resizebox{8.0 cm}{6 cm}{\includegraphics{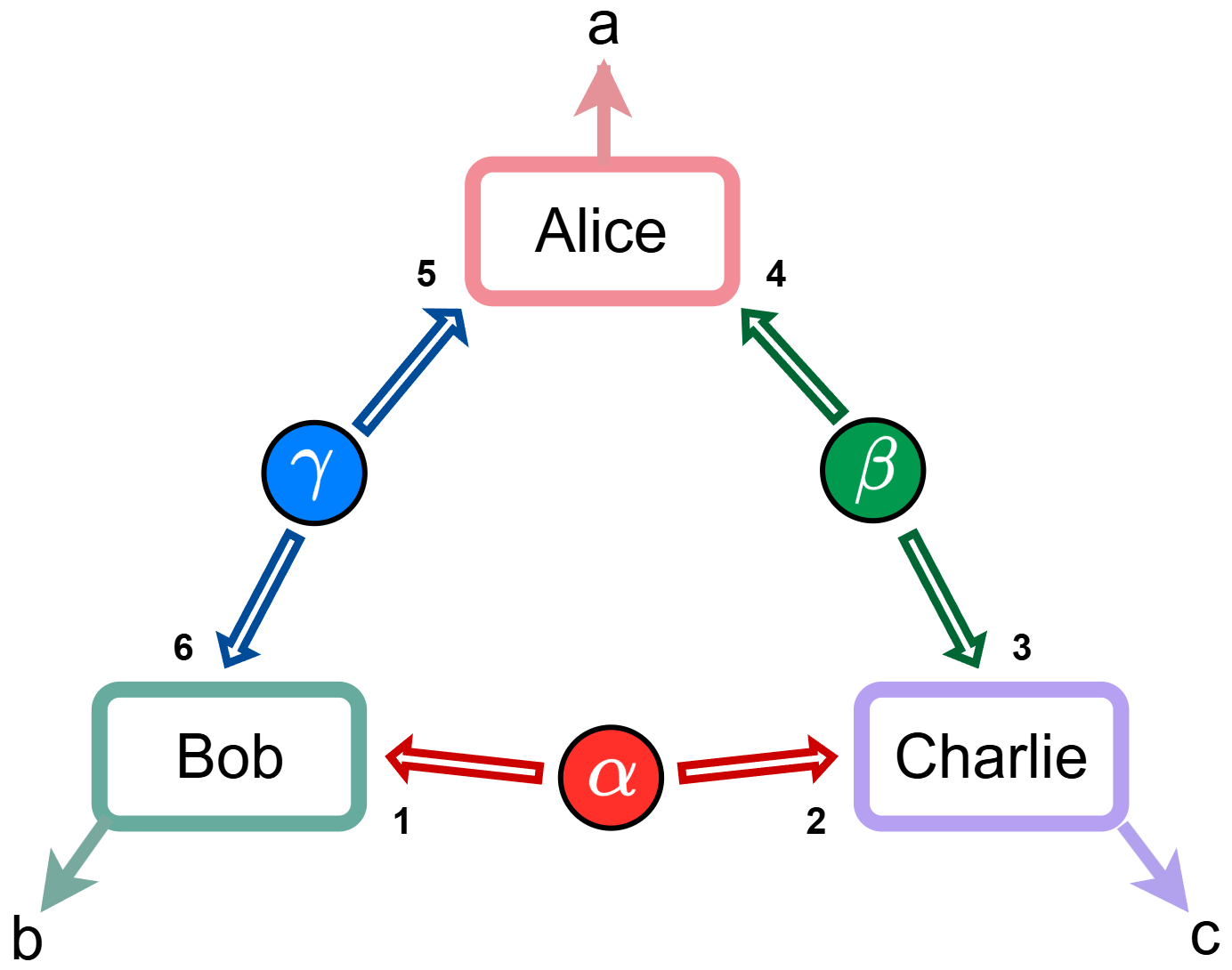}}
  \vspace{-0.15cm}
  \caption{\justifying The triangle network configuration with three sources $\alpha = \rho_{B_1C_2}, \beta = \rho_{C_3A_4} \text{ and } \gamma = \rho_{A_5B_6}$ and three parties (Alice - A, Bob - B, and Charlie - C), each with $4$ measurement outcomes $a,b,c \in \{0,1,2,3\}$.}
  \label{fig:Triangle_network_configuration}
%\vspace{-1.25cm}
\end{figure}

The triangle scenario (see Fig.~\ref{fig:Triangle_network_configuration}) involves three parties in a triangle configuration, Alice, Bob, and Charlie, with three independent sources, where each source sends local variables through a channel to only two of the three parties. Based on the local variables received from the two sources, each party processes its inputs with arbitrary deterministic response functions and outputs a number $a,b,c \in \{0,1,2,3\}$, respectively. The probability distribution $P(a,b,c)$ is obtained from this experiment, and there is a local-realistic hidden variable (LHV) description for the distribution if it has the form:
\begin{equation}
\label{eq:network_trilocal_model}
    p(a,\!b,\!c) \! = \!\! \int \!\! d\alpha \!\! \int \!\! d\beta \! \! \int \!\! d\gamma \, P_A(a|\beta,\gamma) \, P_B(b|\gamma,\alpha) \,P_C(c|\alpha,\beta),
\end{equation}
where $\alpha \in X,\ \beta \in Y,\ \text{and} \ \gamma \in Z$ represents the three local variables distributed by each source, and $P_A(a|\beta,\gamma)$, $P_B(b|\gamma,\alpha)$, and $P_C(c|\alpha,\beta)$ represent the arbitrary deterministic response functions for Alice, Bob, and Charlie. Quantum mechanics allows for distributions that do not admit this local-realistic description~(\ref{eq:network_trilocal_model}), and the set of such distributions that can be generated using quantum resources (see Fig.~\ref{fig:Local_and_nonlocal_domain}) is genuinely network nonlocal.

In the triangle scenario, existing work has primarily focused on symmetric scenarios with each source producing an identical, pure, maximally entangled Bell state, e.g.,
$$|\psi_{\gamma}\rangle_{A_{\gamma}B_{\gamma}} = |\psi_{\alpha}\rangle_{B_{\alpha}C_{\alpha}} = |\psi_{\beta}\rangle_{C_{\beta}A_{\beta}} = \frac{1}{\sqrt{2}}(|00\rangle + |11\rangle),$$ 
with each party performing a projective quantum measurement in the same basis, given by:
\begin{align} \label{eq:uw_povm}
M_0 = u|00\rangle + (\sqrt{1-u^{2}})|11\rangle, \ \ \ & \ \ \ M_2 = |01\rangle\nonumber,\\
M_1 = (\sqrt{1-u^{2}})|00\rangle - u|11\rangle, \ \ \ & \ \ \ M_3 = |10\rangle, 
\end{align}
with $u^2 \in [0.5,1]$.

Contrary to the standard Bell nonlocality tests, where a referee provides Alice and Bob with two sets of measurement choices, here the observers receive no external inputs and instead perform a fixed joint entangled measurement on their systems. Of the three Bell assumptions, the assumption of measurement choice is replaced with source independence in the case of networks. The statistics of the experiment are then described by the resulting joint probability distribution $p(a,b,c)$, where
\begin{align}
    \label{eq:Pq_distribution}
    p(a,b,c) = |\langle M_a|\langle M_b| \langle M_c| |\psi_{\gamma}\rangle |\psi_{\alpha}\rangle |\psi_{\beta}\rangle|^2.
\end{align}

Previous studies have proved that genuine network non-local correlations can be exhibited in the range $0.785<u^2<1$ while an LHV description can be proven to exist for the cases of $u^2=0.5 \ \text{and} \ 1.0$. This specific set of distributions is called the RGB4 \cite{Renou_Genuine2019:PhysRevLett.123.140401} family of distributions. Multidisciplinary approaches using machine learning (ML), such as local hidden variable (LHV-Net) network models, have shown the existence of GNN correlations in the range $u^2 \in \{0.63,0.85\}$ \cite{Tamas_2020}. 

However, all these approaches aimed at demonstrating the existence of genuine network nonlocality struggle with distributions arising from mixed states and exhibit ambiguity in such cases. As a result, they can only offer presumptive estimates on the noise robustness of genuine network nonlocality. Fundamentally, the existing approaches are not equipped to characterize the full range of quantum correlations, particularly those contained in mixed states. This highlights the absence of a robust toolkit for rigorously exploring genuine network nonlocality beyond idealized resources, which is a gap our work directly addresses.

\begin{figure*}[t]
    \centering

    \begin{tikzpicture}

    \node (a) at (0,0)
      {\includegraphics[width=2\columnwidth]{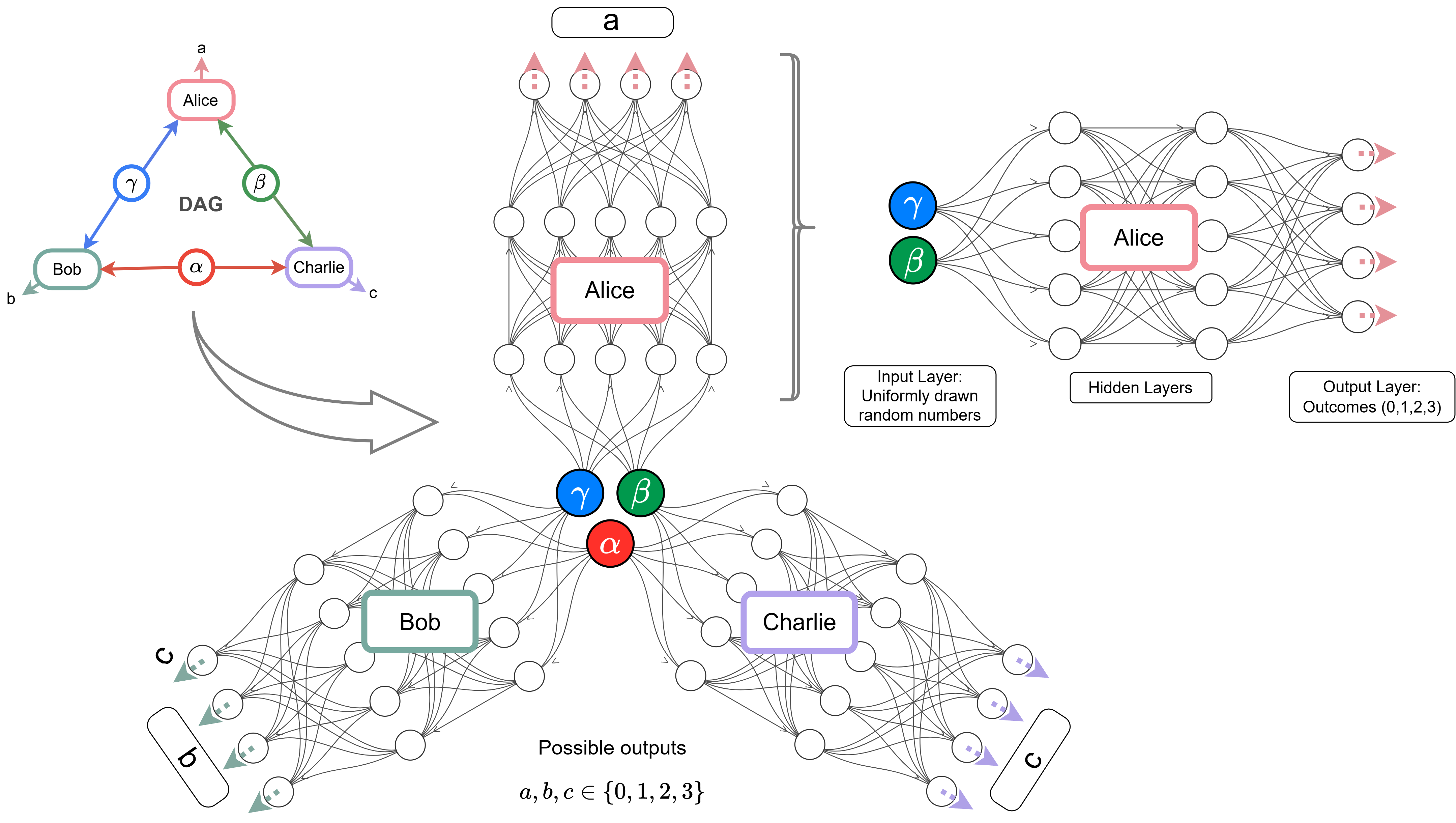}};
        
    \node at (-6.1,-0.3) {\scriptsize{\textbf{(a)}}};
    \node at (-1.4,-5) {\scriptsize{\textbf{(b)}}};
    \node at (5,-0.3) {\scriptsize{\textbf{(c)}}};
    
    \end{tikzpicture}
    \vspace{-0.25cm}
    \caption{\justifying LHV neural network with single-layer response function: (a) The Directed Acyclic Graph (DAG) representation for the triangle network scenario, (b) the local hidden variable (LHV) neural network model following the triangle network scenario's DAG with randomly generated classical numbers as sources, (c) the single-layer response function of each observer represented by a feedforward neural network.}
    \label{fig:System_A}
    \vspace{-0.25cm}
\end{figure*}

\section{Methodology}

To tackle non-convex optimization and cumbersome non-ideal Bell scenarios involving mixed states and noise, we employ machine learning as a favorable alternative, raising it as a quantum framework by embedding quantum foundations and causal inference. We use causally constrained Bayesian networks, which are Directed Acyclic Graphs (DAGs) where each node represents a conditional probability distribution. By preserving the DAGs, a feedforward neural network can then replace the conditional distribution with neural network layers that model it. The graph structure, or the causal constraints of the Bell scenario, remains fixed, but the functions, or strategies, can now be learned. In the triangle network scenario, the standard causal Bayesian network for local hidden variables (LHV) is 
\begin{itemize}
    \item Hidden nodes $\lambda_{1}$, $\lambda_{2}$, and $\lambda_{3}$ (independent sources).
    \item Outputs $a, b, c \in \{0,1,2,3\}$ (measurement outcomes).
\end{itemize}
The conditional independence constraint is given by:
\begin{align}
    p(a,b,c|\lambda_{1}\lambda_{2}\lambda_{3}) = p(a|\lambda_{2},\lambda_{3})p(b|\lambda_{1},\lambda_{3})p(c|\lambda_{1},\lambda_{2}).
\end{align}

By relying on universal approximation \cite{Deep_Learning:Goodfellow-et-al-2016}, a sufficiently expressive model respecting the causal probability relations of the network topology can approximate the local distributions of the Bell scenario in its inner structure. By embedding learning algorithms like Bayesian networks in the framework, although algorithmic in nature, we answer the question of whether the Bell test statistics can be learned or not. Crucially, demonstrating that answering this using a local model is equivalent to certifying it as local. This fundamental difference raises the utility of the model from a data-driven model to a foundational framework.

\subsection{LHV Neural Network Oracle with Single-Layer Response Function}

The RGB4 distributions \cite{Renou_Genuine2019:PhysRevLett.123.140401}, being genuine network non-local, cannot be represented by any LHV description of the form given in Eq.~(\ref{eq:network_trilocal_model}).
Following the DAG of the triangle scenario, a Bayesian network (see Fig.~\ref{eq:network_trilocal_model}) can be developed for the triangle network scenario as an LHV neural network architecture. Here we assume without loss of generality that the three independent sources on the triangle each send a random variable drawn from a uniform distribution on the continuous interval between $0$ and $1$, respectively. This classical neural network representation conserves the constraints of the physical quantum network system capable of only network local distributions; therefore, the joint probability distribution over the party’s outputs can be written in the form given in Eq.~(\ref{eq:network_trilocal_model}).

If the target distribution is local, then the neural network, provided it is sufficiently expressive (trained), will learn the approximate response functions according to the universal approximation theorem (up to an error level). For distributions outside the local set, the machine fails to approximate the given targets; this, in turn, provides a criterion for characterizing distributions of causal structures outside of an LHV description. The LHV-Net model proposed by Tamas et al. \cite{Tamas_2020} is such a neural network oracle, with each party in the network modeled by a feed-forward perceptron-based neural network that attempts to learn the given target distribution over an observed set of variables.

\begin{figure}[t!]
\centering
\begin{tikzpicture}

\node (a) at (0.5,0)
  {\includegraphics[width=0.8\columnwidth]{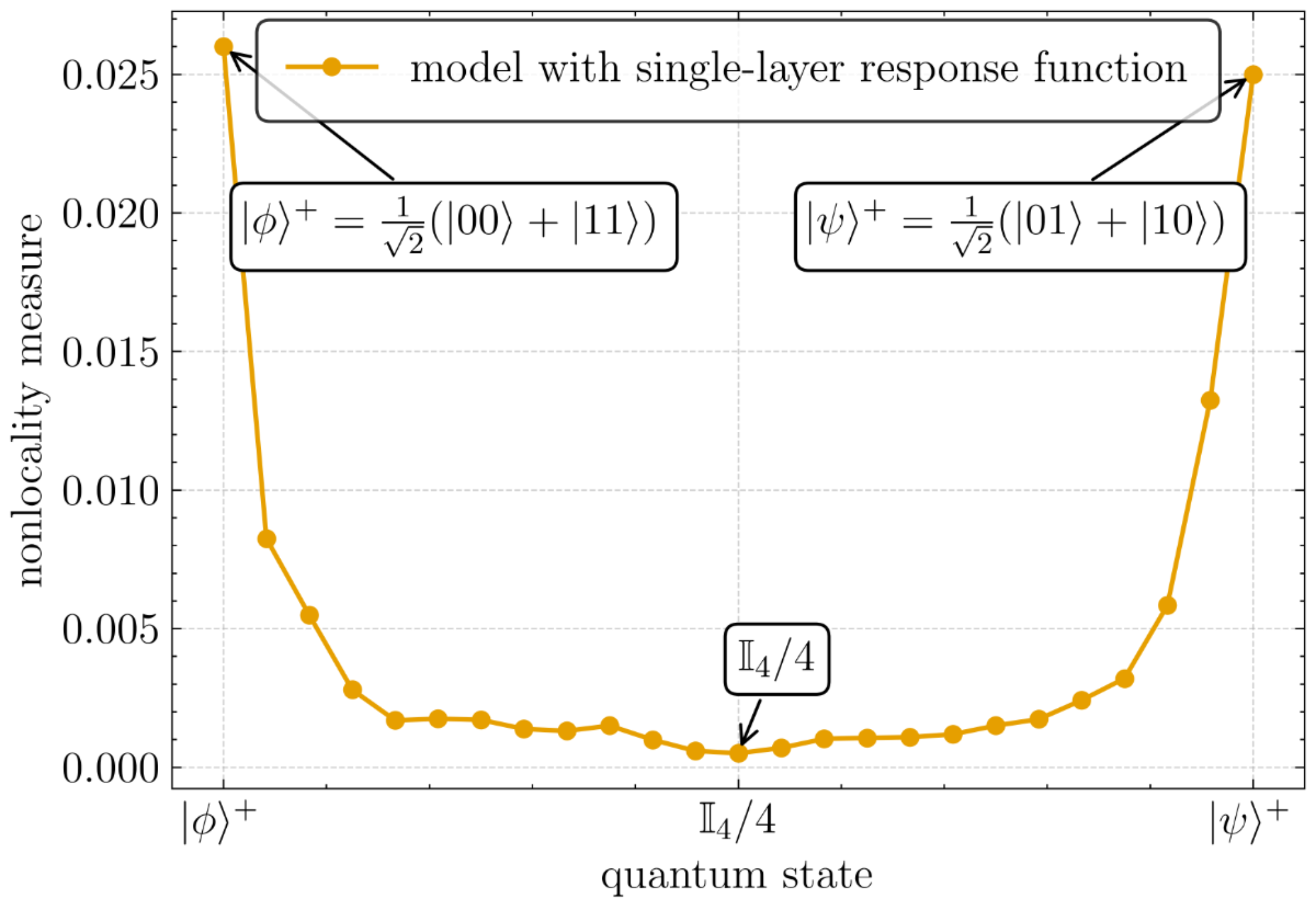}};

\node (b) at (0.5,-0.6\columnwidth)
  {\includegraphics[width=0.8\columnwidth]{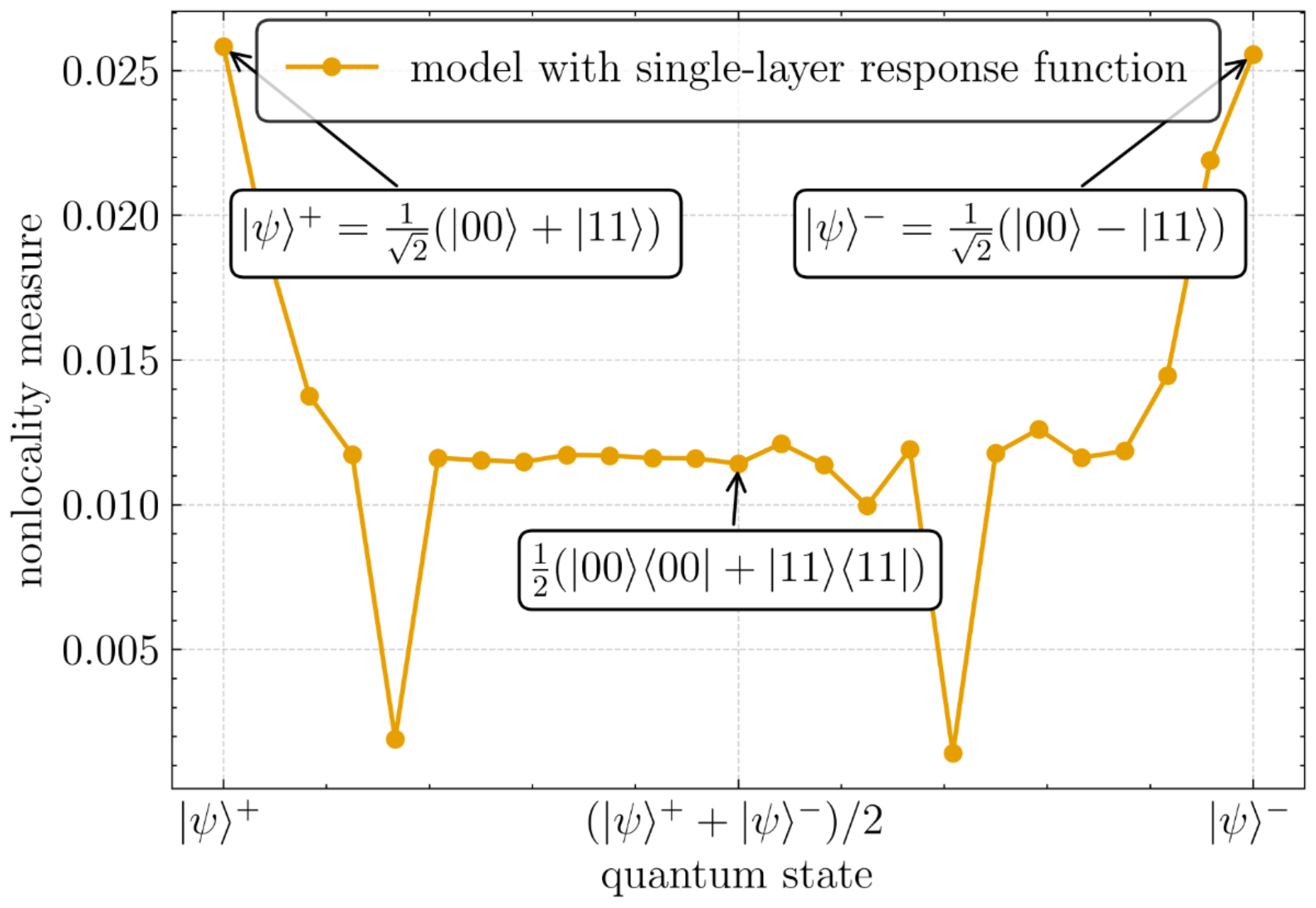}};

\node at (-2.8,-2.33) {\scriptsize{\textbf{(a)}}};
\node at (-2.8,-7.47) {\scriptsize{\textbf{(b)}}};

%\node at ($(a.north west)+(0.15,-0.15)$) {\textbf{(a)}};
%\node at ($(b.north west)+(0.15,-0.15)$) {\textbf{(b)}};

\end{tikzpicture}
\vspace{-0.25cm}
\caption{\justifying Model optimization bottlenecks when characterizing mixed-state distributions with existing models using single-layer response functions. (a) The model fails to give a clear expression for distributions from mixed states that closely match Bell states. (b) The model fails to learn the LHV description of distributions from classically correlated states $\psi_{B_1C_2} = \psi_{C_3A_4} = \psi_{A_5B_6} = \frac{1}{2}(|00\rangle\langle00| + |11\rangle\langle11|)$ in the triangle network. This makes studies on genuine network nonlocality for non-ideal scenarios ambiguous.}
\label{fig:LHV_Model_limitations}
\vspace{-0.25cm}
\end{figure}

Though this method works well with pure states, its limitations arise when dealing with distributions of mixed states in the triangle scenario.  Specifically, for certain classes of classically correlated states and entangled measurements, the model fails to find a local hidden variable expression for genuine network nonlocality. Unlike pure entangled states, analytically proven to express genuine network nonlocality, the model presents ambiguous expressions for mixed-state distributions. These drawbacks are shown in Fig.~\ref{fig:LHV_Model_limitations}, where the disparity that arises when transitioning to mixed states can be clearly seen. Although the neural network expressiveness changes marginally with heavier training, it is unable to deliver clear and conclusive results on the presence or absence of genuine network nonlocality. 

We identify the core issue with the single-layer response function of the model as lacking the required degrees of freedom to learn mixed-state distributions, resulting in ambiguity. This leaves noise robustness studies nonviable and renders the model not suitable for further ambitious studies. Recent works suggest noise robustness \cite{boreiri2024noiserobustproofsquantumnetwork}, but a clear limit has not been achieved so far, making the study of the nature of correlations from non-ideal scenarios an open problem and a hard one owing to the non-convex nature of these correlations, with GNN correlations being speculated to be arbitrarily close to the local boundary \cite{Renou_Genuine2019:PhysRevLett.123.140401}. We formalize and operationally execute a framework answering these questions, including a detailed analysis on the robustness of the genuineness of these correlations to shared randomness, which was an open problem.

\subsection{Learning using the Layered LHV-Net Oracle with Multi-Layer Response Function}

The causal structure of the triangle quantum network scenario with mixed states is fundamentally different from that of pure states. Learning a mixed-state distribution requires the individual response functions of the observer to have more degrees of freedom, as the qubit systems it sends to either of its observers in the triangle scenario are not identical. This makes inner approximating distributions from pure states numerically different from less symmetric cases of mixed states or the cases with added noise. 

Assuming that in a fully separable local model each source emits a bipartite state whose eigenvectors are product states, the corresponding density matrices admit separable spectral decompositions given by:
\begin{align}
\label{eq:bipartite_separable_form}
    \rho^{a}_{B_1,C_2} &= \sum^{k_{a}}_{i=1}\lambda^{a}_i|\psi_i\rangle\langle\psi_i|_{B_1C_2} %\nonumber \\
     %&= \sum^{k_{a}}_{i=1}\lambda^{a}_i|\psi_i\rangle\langle\psi_i|_{B_1}\otimes|\psi_i\rangle\langle\psi_i|_{C_2} \nonumber \\ 
     = \sum^{k_{a}}_{i=1}\lambda^{a}_i\rho^i_{B_1}\otimes\rho^i_{C_2}, \nonumber \\
     \rho^{b}_{C_3A_4} &= \sum^{k_{b}}_{j=1}\lambda^{b}_j\rho^j_{C_3}\otimes\rho^j_{A_4}, \nonumber \\ 
     \rho^{c}_{A_5B_6} &= \sum^{k_{c}}_{l=1}\lambda^{c}_l\rho^l_{A_5}\otimes\rho^l_{B_6},
\end{align}
where $\{|\psi_i\rangle_{B_1C_2}, |\psi_j\rangle_{C_3A_4}, |\psi_l\rangle_{A_5B_6}\}$ are orthonormal bases for the respective local Hilbert spaces, and $\{\lambda^{a}_i, \lambda^{b}_j, \lambda^{c}_l\}$ are the nonzero eigenvalues. Consequently, $\{k_{a}, k_{b}, k_{c}\}$ denotes the rank of the spectral decomposition of the respective states. By taking the tensor product of the three sources, we get a single six-qubit multipartite quantum state $\rho_{B_1 C_2 C_3 A_4 A_5 B_6}$ like in the standard multipartite Bell scenarios as
%\begin{widetext}
\begin{align}\label{eq:tripartite_separable_form}
\rho_{B_1C_2C_3A_4A_5B_6} &= \rho_{B_1C_2} \otimes \rho_{C_3A_4} \otimes \rho_{A_5B_6}\nonumber \\
&= \sum_{i=1}^{k_{a}} \sum_{j=1}^{k_{b}} \sum_{l=1}^{k_{c}} \lambda^{\alpha}_i\lambda^{\beta}_j\lambda^{\gamma}_l \big(\rho_{B_1}^i \otimes \rho_{C_2}^i\big) \nonumber \\
& \qquad  \otimes \big(\rho_{C_3}^j \otimes \rho_{A_4}^j\big) \otimes \big(\rho_{A_5}^l \otimes \rho_{B_6}^l\big).
\end{align}
%\end{widetext}

Considering the triangle network scenario and its transformation on the total quantum state (see Fig.~\ref{fig:Triangle_network_configuration}), we find $\rho_{A_5A_4B_1B_6C_3C_2}$ from swapping the basis of~(\ref{eq:tripartite_separable_form}) to match that of the observer's basis in the network and obtain
%\begin{widetext}
    \begin{align}\label{eq:mixed_separable_state}
    \rho_{A_5A_4B_1B_6C_3C_2} &= \sum_{i=1}^{k_{a}} \sum_{j=1}^{k_{b}} \sum_{l=1}^{k_{c}} \lambda^{\alpha}_i\lambda^{\beta}_j\lambda^{\gamma}_l (\rho_{A_5}^j \otimes \rho_{A_4}^l) \nonumber \\
    & \qquad \quad \otimes (\rho_{B_1}^i \otimes \rho_{B_6}^j) \otimes (\rho_{C_2}^i \otimes \rho_{C_3}^j)\nonumber \\
    &= \sum_{i=1}^{k_{a}} \sum_{j=1}^{k_{b}} \sum_{l=1}^{k_{c}} \lambda^{\alpha}_i\lambda^{\beta}_j\lambda^{\gamma}_l \Big(\rho_{A_5A_4}^{(j,l)_{\rm{sep}}} \nonumber \\
    &\qquad \qquad \qquad  \otimes \rho_{B_1B_6}^{(i,l)_{\rm{sep}}} \ \otimes \rho_{C_2C_3}^{(i,j)_{\rm{sep}}}\Big).
    \end{align}
%\end{widetext}

By using a mixed state, we increase the variety of distributions that can be generated by the system. Hence, each source $\alpha$, $\beta$, and $\gamma$ can now send dissimilar qubits to each of their respective observers. Now, as each observer does a joint POVM measurement on their respective pair of qubits received, the final distribution can be found to be the sum of $k_a \times k_b \times k_c$ distributions, with each of the observers having their conditional probability distribution as a sum of $k_a\times k_b$, $k_b\times k_c$, and $k_c\times k_a$ distributions, where $k_a$, $k_b$, and $k_c$ are the ranks of the spectral decompositions of density matrices of the states each source provides. The probability distribution generated can then be represented as:
%\begin{widetext}
\begin{align}\label{eq:mixed_separable_distribution}
    p(a,b,c) 
    &= {\rm Tr}\big[P_{A_5A_4}^a \!\! \otimes \! P_{B_1B_6}^b \!\! \otimes \! P_{C_3C_2}^c \,\rho_{A_5A_4B_1B_6C_3C_2} \big] \nonumber\\
    &= \sum_{i=1}^{k_{a}} \sum_{j=1}^{k_{b}} \sum_{l=1}^{k_{c}} \lambda^{\alpha}_i\lambda^{\beta}_j\lambda^{\gamma}_l \ {\rm Tr}\big[P_{A_5A_4}^a\rho_{A_5A_4}^{j,l} \big] \nonumber \\ & \qquad \quad \times {\rm Tr} \big[P_{B_1B_6}^b\rho_{B_1B_6}^{i,l} \big] {\rm Tr} \big[ P_{C_3C_2}^c\rho_{C_3C_2}^{i,j} \big].
\end{align}
%\end{widetext}
If the statistics from an experiment do not follow the expression above~(\ref{eq:mixed_separable_distribution}), it is nonlocal by non-network standards. This thus introduces the rank of the state as an additional parameter and a previously untapped resource in creating a local model for the target distribution.

\begin{figure}[t]
  \begin{center}
    \includegraphics[scale = 0.3]{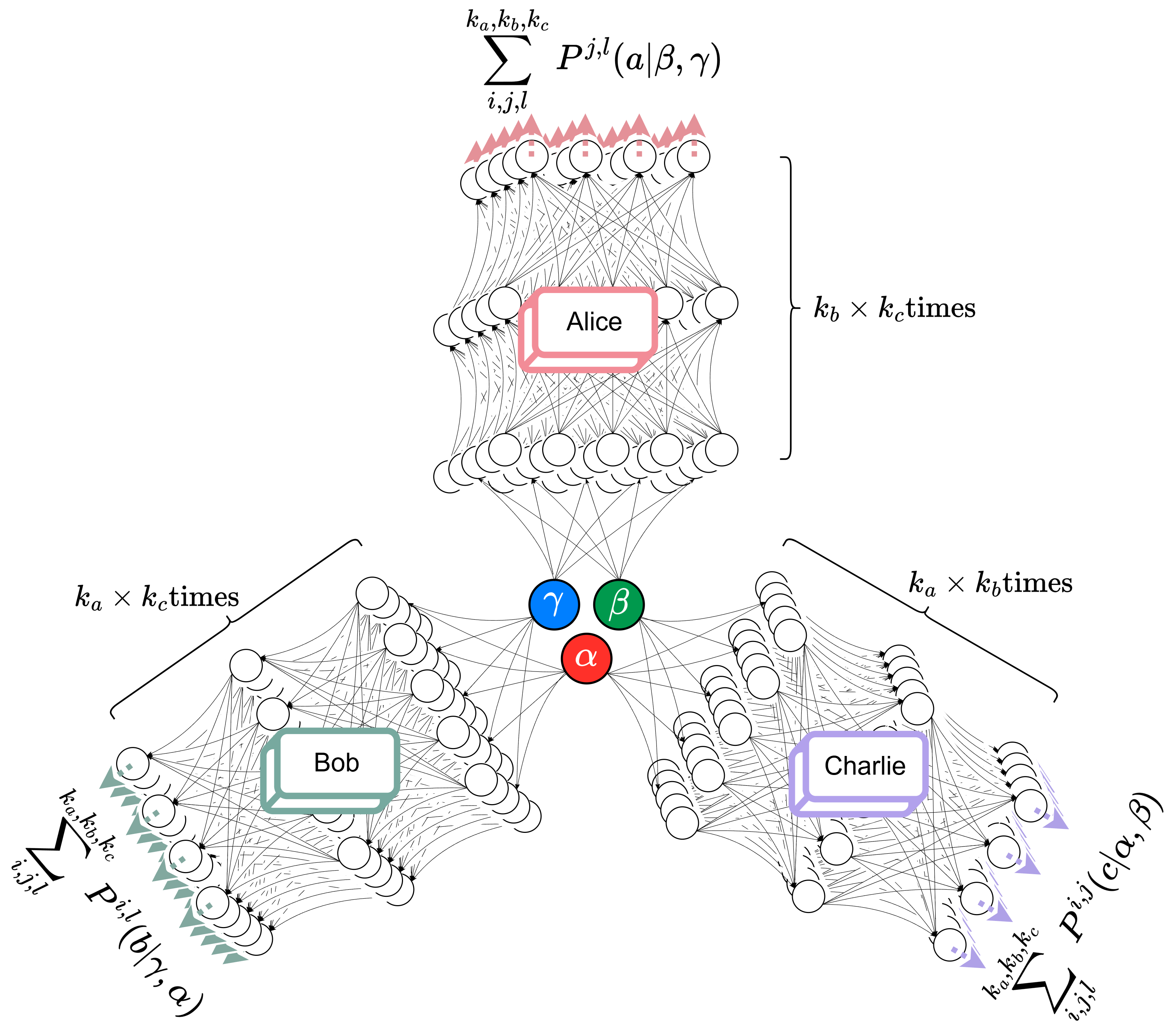}
  \end{center}
  \vspace{-0.25cm}
  \caption{\justifying Layered LHV neural network with multi-layer response functions: Multiple layers of neural networks, each corresponding to the spectral decomposition ranks of the observed qubit subsystems, are trained in parallel to learn the response functions for mixed-state distributions, where $k_a \times k_a$, $k_b \times k_b$ and $k_c \times k_c$ are ranks of the qubit subsystems observed by Alice, Bob, and Charlie, respectively. The response functions for each observers are calculated by summing up from their respectives outcomes $k_a \times k_b \times k_c$ times.}
  \label{fig:System_B}
  \vspace{-0.25cm}
\end{figure}

Using this intuition, we build our numerical analog of the Bayesian network architecture, where we layer the response functions to account for the increased degrees of freedom the information has. The rank of the mixed states informs us about these additional degrees of freedom. By training the layers of neural network response functions of each observer - Alice, Bob, and Charlie in parallel and summing over the $k_a\times k_b\times k_c$ distributions, we can perfectly learn the local description of the target distribution.  It is important to stress here that we are not applying neural networks in their standard setting to learn and predict from data but as a foundational framework using their capability for learning a target distribution or statistics and characterizing it as local based on the neural architecture.

\begin{figure}[t]
    \centering
    \begin{tikzpicture}
    
    \node (a) at (0.5,0)
      {\includegraphics[width=0.8\columnwidth]{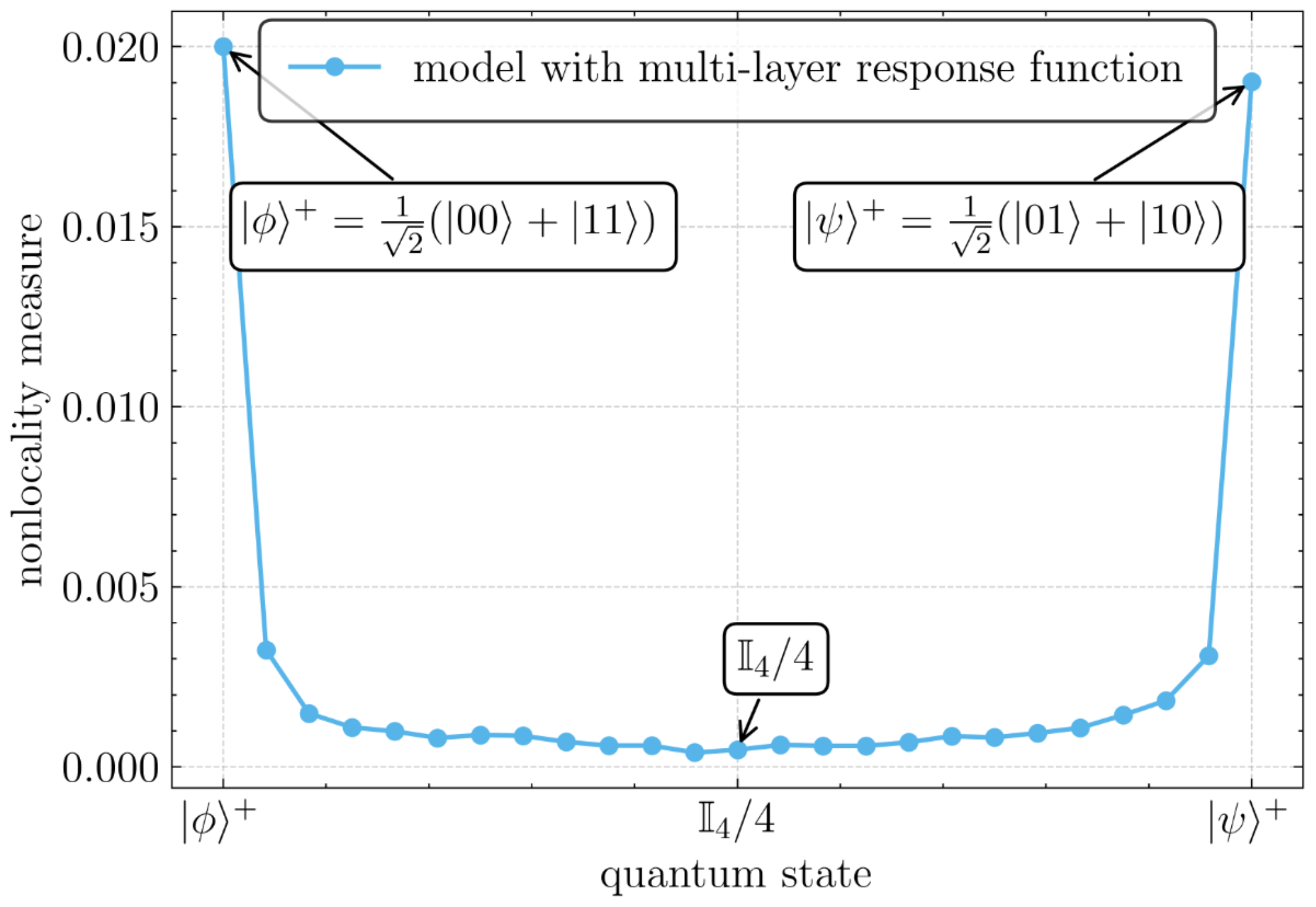}};
    
    \node (b) at (0.5,-0.6\columnwidth)
      {\includegraphics[width=0.8\columnwidth]{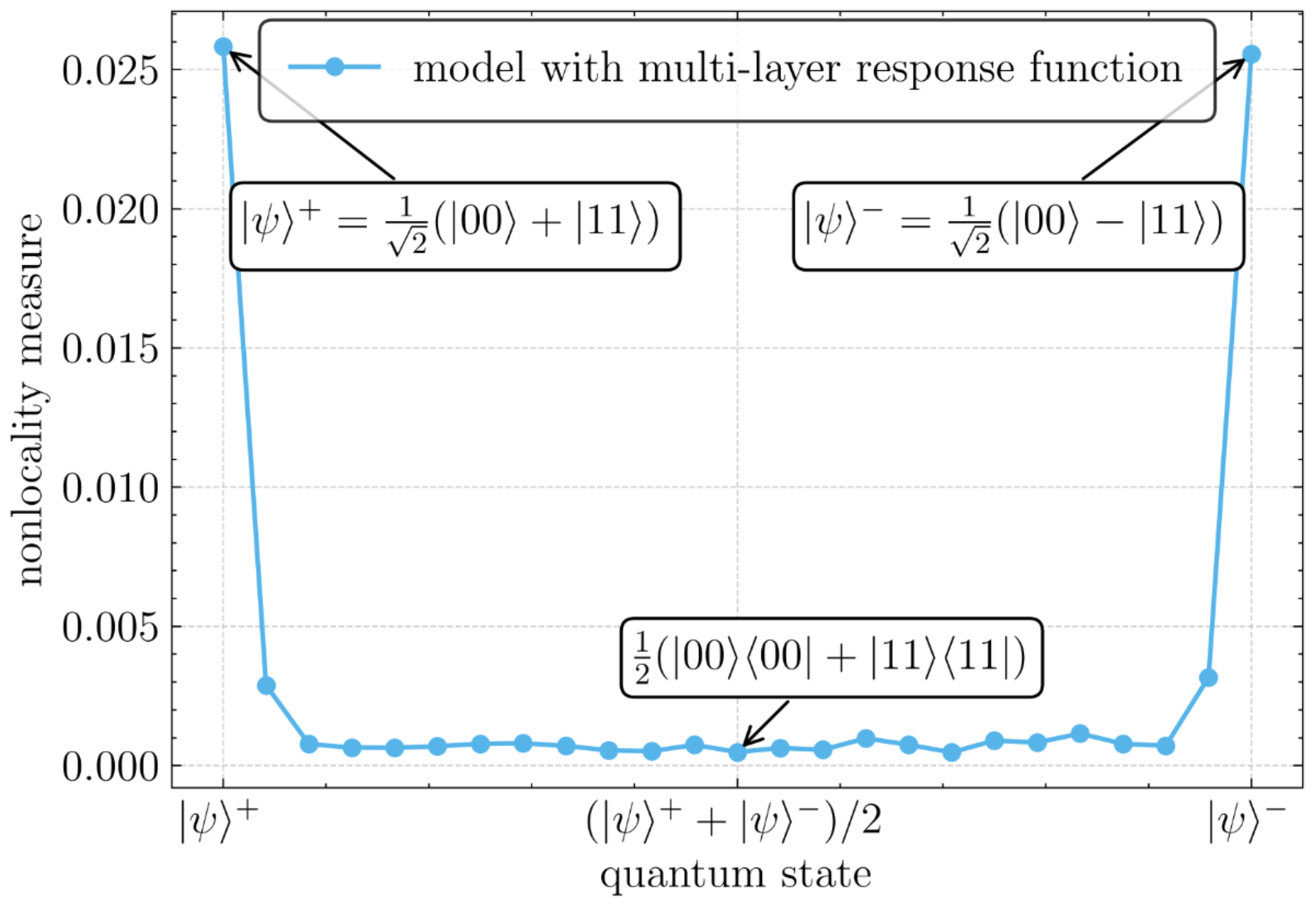}};

    \node at (-2.8,-2.34) {\scriptsize{\textbf{(a)}}};
    \node at (-2.8,-7.47) {\scriptsize{\textbf{(b)}}};
    
    %\node at ($(a.north west)+(0.15,-0.15)$) {\textbf{(a)}};
    %\node at ($(b.north west)+(0.15,-0.15)$) {\textbf{(b)}};
    
    \end{tikzpicture}
    \vspace{-0.25cm}
    \caption{\justifying Making the model response functions layered solves the limitations of characterizing genuine network nonlocality. The new model with the $2\times2$ layered response function gives a clear characterization for mixed-state distributions, including (a) distributions from states that closely match Bell states and (b) the previously ambiguous classically correlated state distributions.}
    \label{fig:LHV_model_correction}
    \vspace{-0.25cm}
\end{figure}

\subsection{Implementation}

We first implement (see Fig.~\ref{fig:System_A}) the framework for pure states, with single-layer response functions for each observer constrained by the DAGs of the network topology, respectively.  Next (see Fig.~\ref{fig:System_B}), we extend the implementation to mixed states with multi-layer response functions. For the triangle network scenario, the three single-layer response functions, or conditional probability functions, are modeled by perceptron neural networks for each of the agents (see Fig.~\ref{fig:System_A}c), Alice, Bob, and Charlie, respectively. These include the ReLU activation layers and a softmax function at the last layer to impose normalization as each observer's probability is being targeted.

Drawing parallels to the physical system we numerically simulate, the inputs are the hidden variables, i.e., uniformly drawn random numbers $\alpha,\,\beta,\,\gamma$. And the outputs are the conditional probabilities $P_{A}(a|\beta,\,\gamma), \, P_{B}(b|\gamma, \alpha)$, and $P_{C}(c|\alpha,\,\beta)$ for Alice, Bob, and Charlie, respectively, three normalized vectors, each of length four. To respect the communication constraints of the triangle systems, the three neural networks are not fully connected with each other, as shown in Fig.~\ref{fig:System_A}b.

Coming to mixed-state distributions, we require as many layers as the rank of the qubit systems being measured for the observer response functions (see Fig.~\ref{fig:System_B}). For the extended implementation, we use an array of layers parallelly stacked for each observer response function connected by their source inputs and response outputs, $k_b \times k_c$ for Alice, $k_a \times k_c$ for Bob, and $k_a \times k_b$ for Charlie, with $k_a$, $k_b$ and $k_c$ being the spectral decomposition rank of the qubit systems that are shared. 

Without loss of generality, we assign random variables drawn from a uniform distribution on the continuous interval between $0$ and $1$ as sources. To train the neural network, we synthetically generate uniform random numbers for the hidden variables, the inputs. We then adjust the weights of the neural network after evaluating the cost function on a batch size, $N_b$, using any standard neural network optimization criterion. For the loss function, we used the Kullback divergence, which measures the discrepancy between the two distributions, given by:
\begin{align}
\label{eq:Kullback_Divergence_loss}
    L(p_m) &= \sum_{a,b,c} \ p_t(a,b,c) \log \bigg[ \frac{p_t(a,b,c)}{p_m(a,b,c)} \bigg],
\end{align}
where $p_t(a,b,c)$ is the target distribution we want to learn, and $p_m(a,b,c)$ is the learned distribution by the model. For every scenario, we evaluate the neural network for $N_{batch}$ values of $\alpha,\,\beta,\,\gamma$ as the inputs for the model to approximate the joint probability distribution with a Monte Carlo approximation, where
%\begin{widetext}
\begin{align}
\label{eq:Monte_carlo_approximation}
    p_{N_{batch}}(a,b,c) \! &=  \!\! \frac{1}{N_{b}} \!\! \sum^{N_{b}}_{n=1} \! \sum_{i=1}^{k_{a}} \! \sum_{j=1}^{k_{b}} \! \sum_{l=1}^{k_{c}} \!\lambda^{\alpha}_i \lambda^{\beta}_j \lambda^{\gamma}_l {\rm Tr} \big(P_{A_5A_4}^a\rho_{A_5A_4}^{j,l}\big)\nonumber \\
    & \quad  \times \, {\rm Tr} \big(P_{B_1B_6}^b \rho_{B_1B_6}^{i,l}\big) {\rm Tr}\big(P_{C_3C_2}^c\rho_{C_3C_2}^{i,j}\big).
\end{align}    
%\end{widetext}

The training is done in parallel on all separate layers of the observers, such that there are $k_a\times k_b$, $k_b\times k_c$, and $k_a \times k_c$ distributions for Alice, Bob, and Charlie, respectively, summed over $k_a \times k_b \times k_c$ times approximating the target distribution, $k_a$, $k_b$ and $k_c$ being the rank of the state density matrices. Additionally, we allow the neural network to absorb the coefficients in the separable decomposition as weights into its trainable parameters, instead of explicitly parameterizing them. The products $\lambda^{\alpha}_i \lambda^{\beta}_j \lambda^{\gamma}_l$ are implicitly represented and adjusted during optimization, and their effective contribution is determined by the optimization of the loss function.

This neural network architecture, introducing the rank of the qubit subsystems under measurement as a parameter, allows the simulation of the network scenario by adapting the required degrees of freedom of the observers. This framework demonstrates a marked improvement over existing techniques in addressing the robustness of the causal inference task while overcoming the earlier limitation (see Fig.~\ref{fig:LHV_model_correction}) in studying distributions with mixed states. It achieves high accuracy in determining the existence of a local model, with significantly reduced computational cost. Furthermore, this paves the way for interesting results when applied to the triangle network scenario and when studying its noise robustness, followed further by other studies on the genuineness of the correlations, which wouldn't have been viable using earlier methods.

We characterize the target distribution using a Euclidean distance measure between the target and model distributions, 
\begin{align}
\label{eq:distance_measure}
    d(p_t,p_m) = \sqrt{\sum_{a,b,c}[p_t(a,b,c) - p_m(a,b,c)]^2}.
\end{align}
After optimization or learning, the measure should reduce to a very small error factor if the target distribution is local. In the output statistics, observing a clear increase in distance $d(p_t,p_m)$ at some point signals that the distribution is leaving the local set.

While we use the rank of the states as input to make the framework, our algorithm, derived from the framework, remains operational irrespective of the actual rank, as it does not place any requirements on the state or measurement to characterize the statistics of the experiment. We therefore employ the $2\times2$ layered response functions for our study, which has the necessary degrees of freedom/resources compared to the $1\times1$ response function to simulate mixed-state distributions faithfully while remaining practically feasible compared to the high computational cost of $3\times3$ and $4\times4$ layered response functions.

\begin{figure*}[t]
\centering
\begin{tikzpicture}

\node (a) at (0.5,0)
  {\includegraphics[width=1\columnwidth]{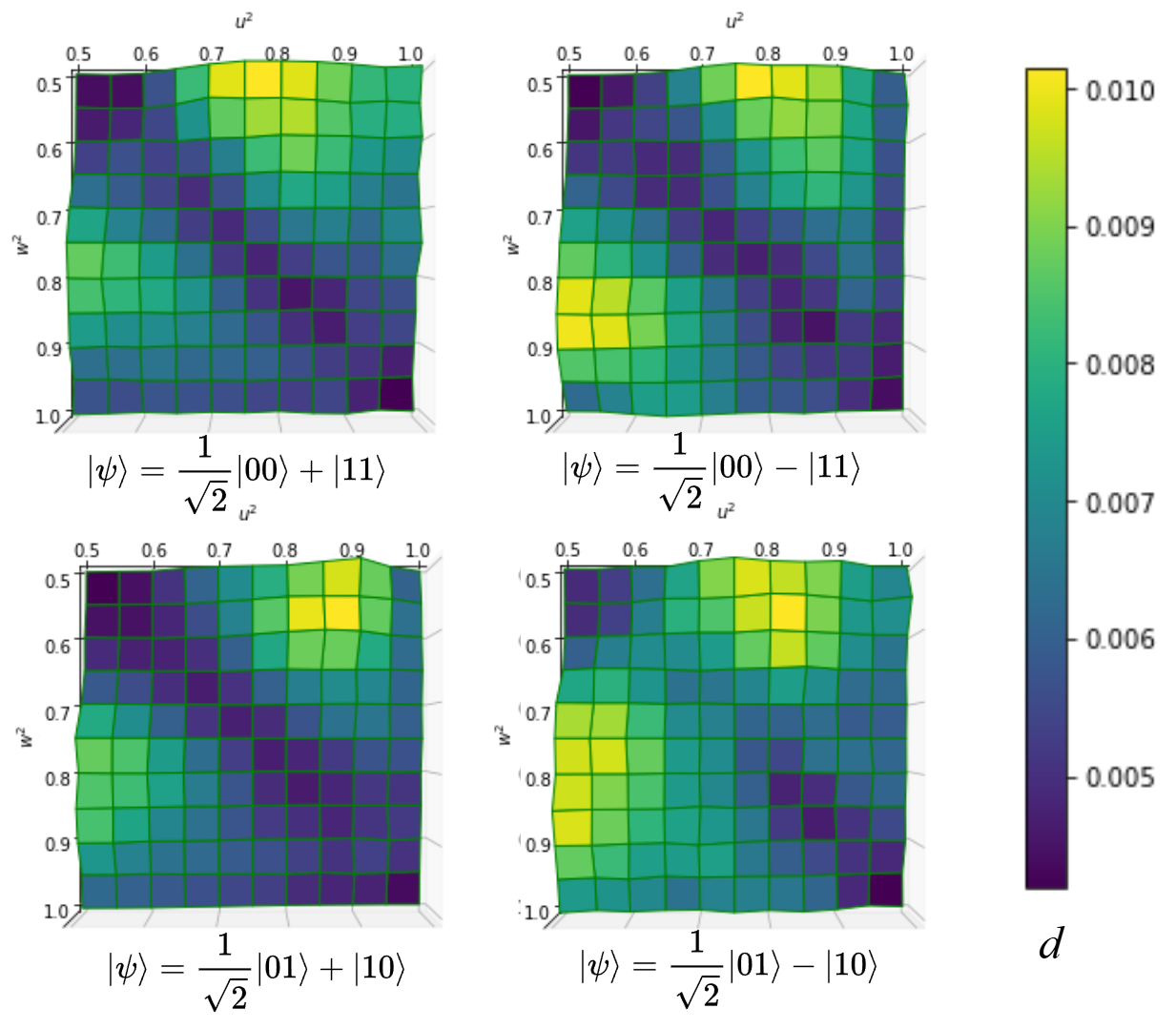}};

\node (b) at (\columnwidth,-0.15)
  {\includegraphics[width=1.05\columnwidth]{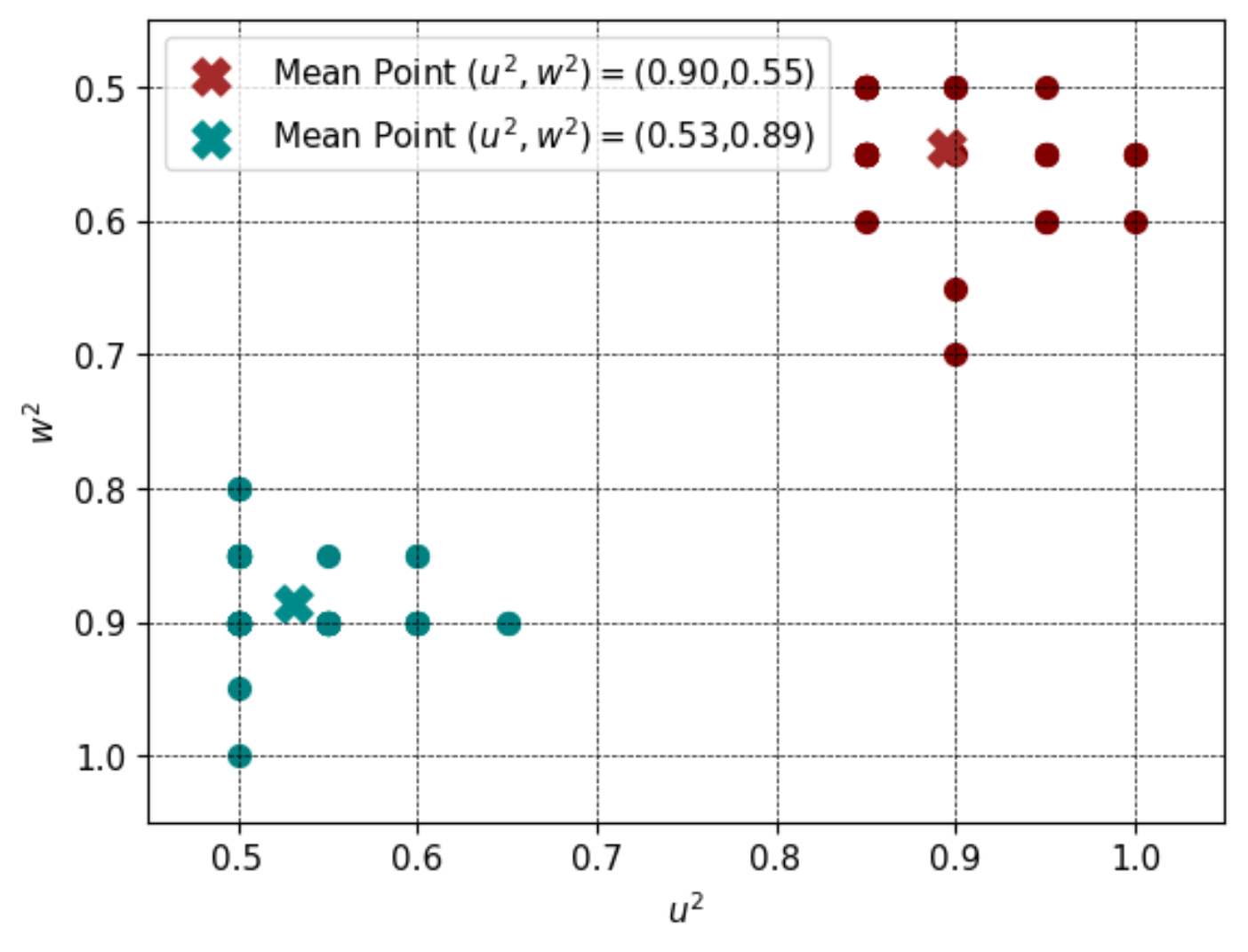}};

\node at (0,-4.0) {\scriptsize{\textbf{(a)}}};
\node at (9.1,-4.0) {\scriptsize{\textbf{(b)}}};

%\node at ($(a.north west)+(0.15,-0.15)$) {\textbf{(a)}};
%\node at ($(b.north west)+(0.15,-0.15)$) {\textbf{(b)}};

\end{tikzpicture}
\vspace{-0.35cm}
\caption{\justifying (a) Genuine network nonlocality (GNN) expression for the four Bell states, showing two symmetric maxima in measurement parameters $(u^2,w^2)$. The measure $d$ denotes the Euclidean distance of the target distribution from the local set. (b) Optimal measurement settings averaged over all states, with optimal values yielding maximal network nonlocality at $(u^2,w^2) = (0.875,0.550)$ followed by $(0.550,0.875)$.}
\label{fig:box1}
\vspace{-0.25cm}
\end{figure*}

\begin{figure*}[t]
\centering
\begin{tikzpicture}

\node (a) at (0,0)
  {\includegraphics[width=0.9\columnwidth]{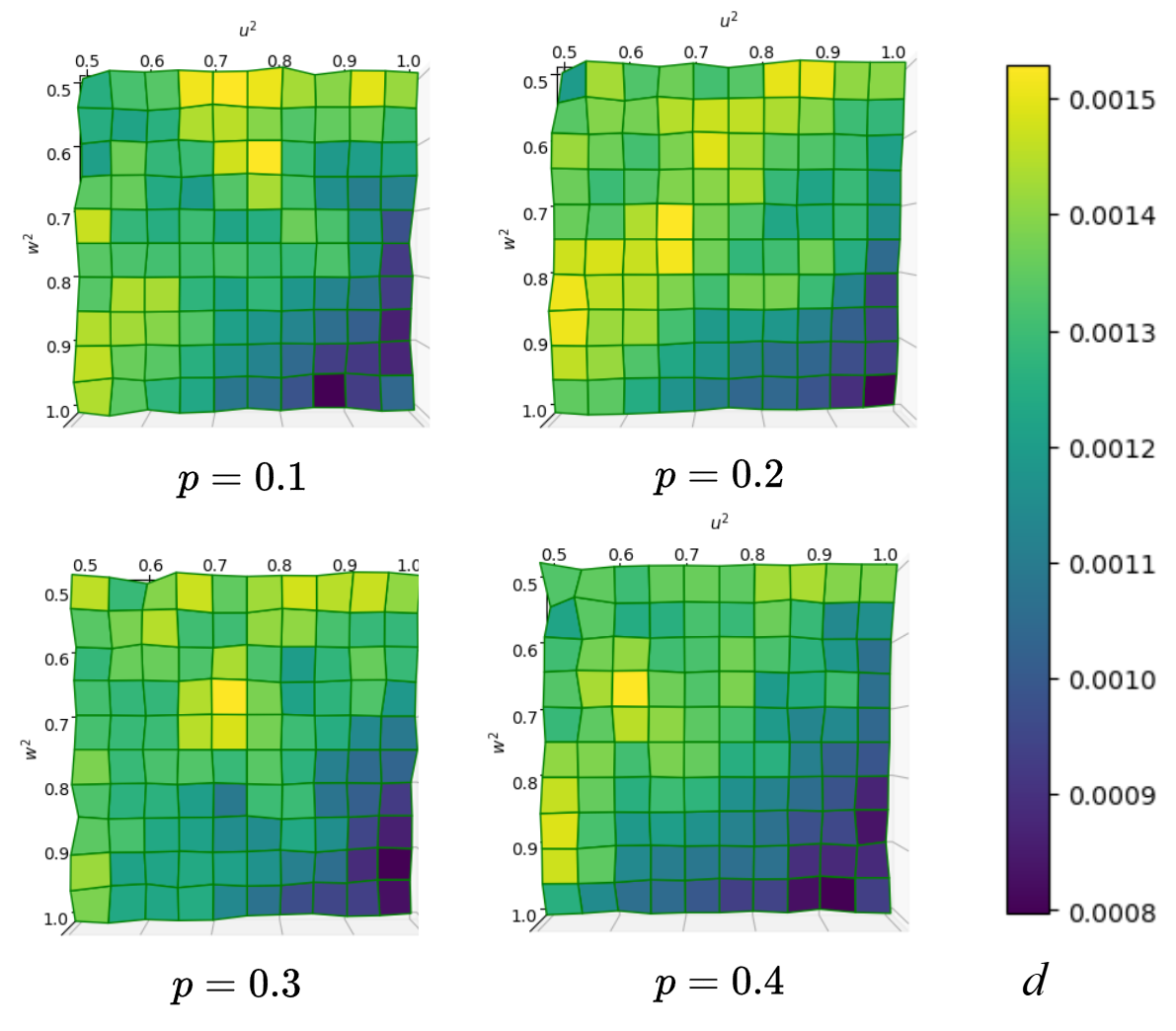}};

\node (b) at (\columnwidth,-0.15)
  {\includegraphics[width=1.1\columnwidth]{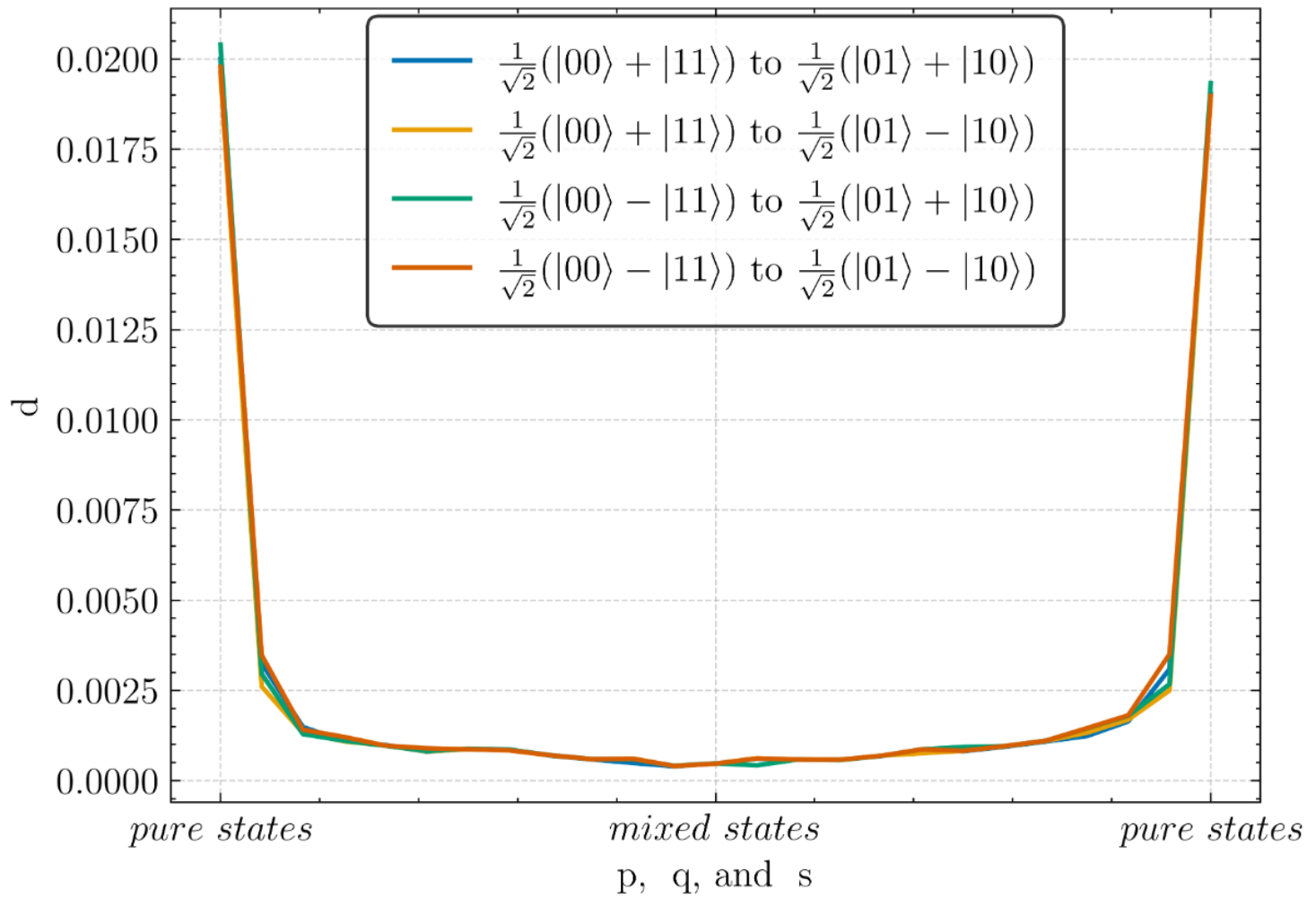}};

\node at (0,-3.8) {\scriptsize{\textbf{(a)}}};
\node at (9.1,-3.8) {\scriptsize{\textbf{(b)}}};

%\node at ($(a.north west)+(0.15,-0.15)$) {\textbf{(a)}};
%\node at ($(b.north west)+(0.15,-0.15)$) {\textbf{(b)}};

\end{tikzpicture}
\vspace{-0.35cm}
\caption{\justifying (a) Genuine network nonlocality (GNN) expression for mixed-state distributions parameterized by $p,q,s$ (trace unity $r=0.5-p$) with no violation observed in the bulk of the parameter space. The measure d denotes the Euclidean distance from the local set. (b) Transitioning between pure and mixed-state distributions, the GNN expression attains its maximum near pure states and vanishes towards the interior of the mixed-state distributions. The optimal measurement settings remain $(u^2,w^2) = (0.550,0.875)$ and $(0.875,0.550)$.}
\label{fig:box2}
\vspace{-0.25cm}
\end{figure*}

\begin{figure*}[t]
\centering
\begin{tikzpicture}

\node (a) at (0,0)
  {\includegraphics[width=0.95\columnwidth]{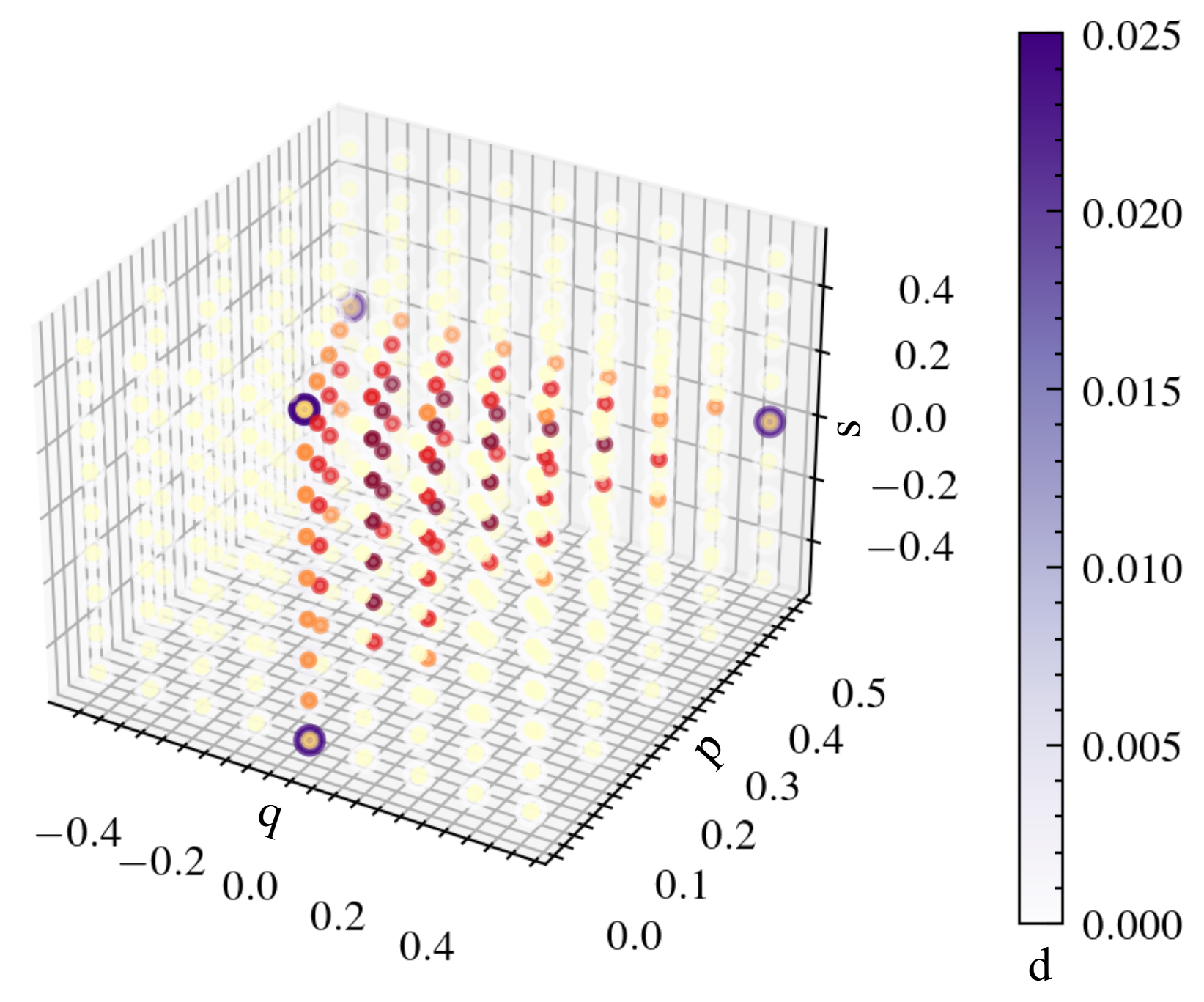}};

\node (b) at (\columnwidth,-0.15)
  {\includegraphics[width=1.05\columnwidth]{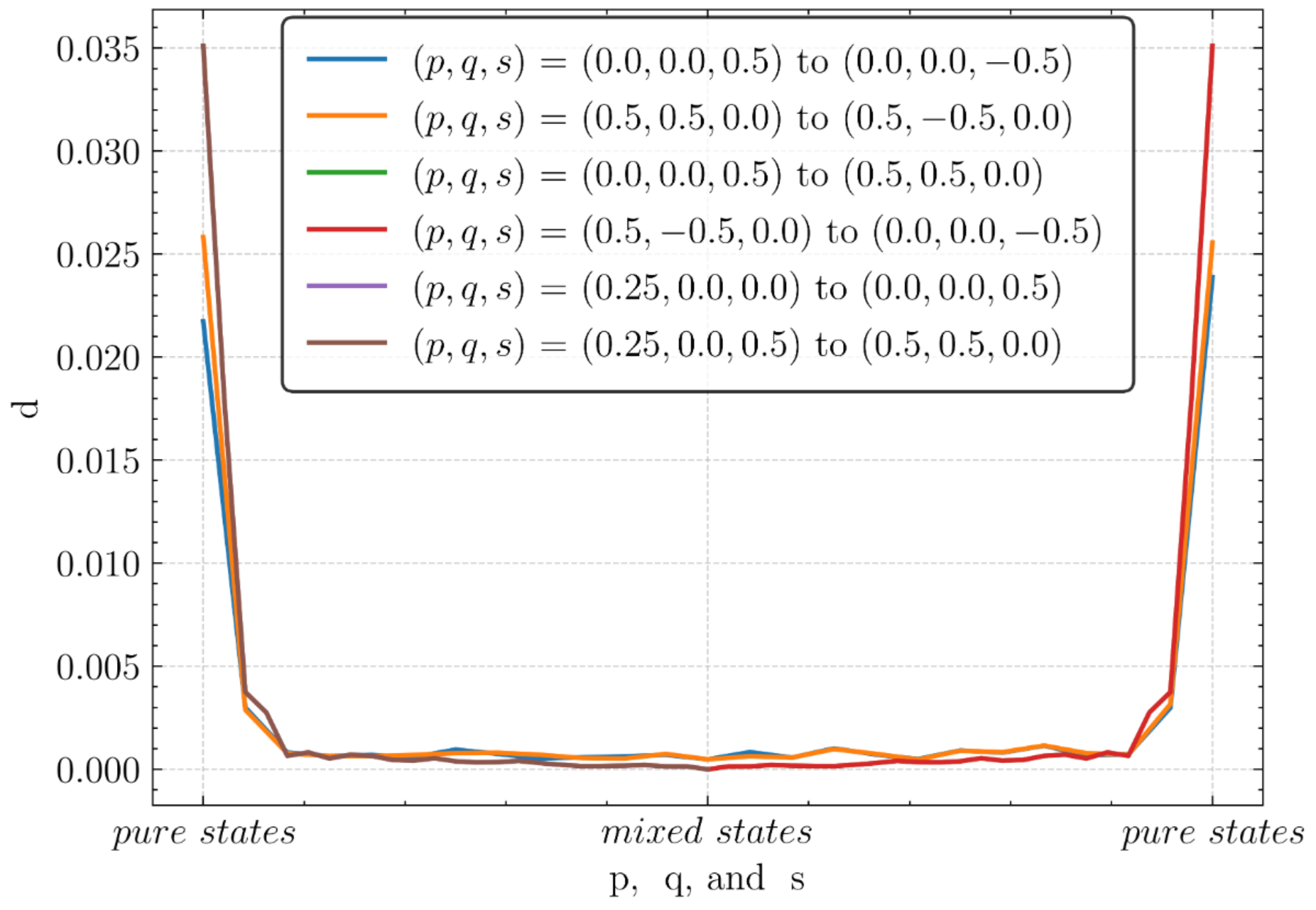}};

\node at (0,-3.8) {\scriptsize{\textbf{(a)}}};
\node at (9.1,-3.8) {\scriptsize{\textbf{(b)}}};

%\node at ($(a.north west)+(0.15,-0.15)$) {\textbf{(a)}};
%\node at ($(b.north west)+(0.15,-0.15)$) {\textbf{(b)}};

\end{tikzpicture}
\vspace{-0.35cm}
\caption{\justifying (a) We consider a general class of two-qubit states whose physically valid states form a tetrahedron in the parameter space $(p,q,s)$, where the four vertices correspond to pure states and the interior represents mixed states. (b) The GNN expression, quantified via the Euclidean distance $d$, is concentrated near the pure-state vertices of the tetrahedron and diminishes toward the interior. The optimal measurement setting being $(u^2,w^2) = (0.550,0.875)$ \& $(0.875,0.550)$.}
\label{fig:tetrahedron}
\vspace{-0.25cm}
\end{figure*}

\section{Results}

We focus on Bell-like states that can exhibit genuine network nonlocality in the triangle network scenario. Specifically, we look at a class of mixed states that span between the maximally mixed state and certified cases of entangled Bell states. But before looking at mixed states, we first present our results by exploring the case of pure states and benchmarking the best measurement setting for maximal genuine network nonlocality expression. 

We also study the noise robustness of the correlations with respect to Werner noise, and we find a new limit to Werner noise robustness, establishing their close vicinity to the local boundary, further benchmarking the utility of our framework. We further explore the triangle network scenario with sources subject to dissimilar Werner noise values. We then study distributions with dissimilar levels of entanglement to identify whether all sources require entanglement in order to express genuine network nonlocality. Finally, we check the robustness of genuine network nonlocality to shared randomness with all sources and partial shared randomness with any two sources.

%\subsection{Genuine Network non locality of \texorpdfstring{$X$}{X} Mixed States}
\subsection{Genuine Network Nonlocality for a Class of Mixed States}

For studying the nature of genuine network nonlocal correlations of mixed-state distributions, we use so-called, mixed `$X$-states' of the form: \vspace{-0.2cm}\begin{align}
\label{eq:pqs_matrix}
    \rho^{\alpha}_{B_{\alpha}C_{\alpha}} = \rho^{\beta}_{C_{\beta}A_{\beta}} = 
    \rho^{\gamma}_{A_{\gamma}B_{\gamma}} =  
    \begin{bmatrix}
    p & 0 & 0 & q\\
    0 & r & s & 0\\
    0 & s & r & 0\\
    q & 0 & 0 & p
    \end{bmatrix},
\end{align}
where $r = 0.5-p$. For generating distributions in the triangle scenario, we take these three effective parameters $p,q,$ and $s$ that traverse the set of mixed states by mapping over the Bell states $|\phi^+\rangle = (|00\rangle + |11\rangle)/\sqrt{2}$, $|\phi^-\rangle = (|00\rangle - |11\rangle)/\sqrt{2}$, $|\psi^+\rangle = (|01\rangle + |10\rangle)/\sqrt{2}$, $|\psi^-\rangle = (|01\rangle - |10\rangle)/\sqrt{2}$ and their diagonal elements. 

We focus on $X$ states for several reasons. Realistic network resources are seldom pure, and $X$ states retain the essential symmetry of networks accommodating these imperfections, making them more practical for studying genuine network nonlocality. Secondly, when measurements are non-maximal, mixed states often match the measurement asymmetry better than pure maximally entangled states \cite{Optimal_Teleportation_with_a_Mixed_State_of_Two_Qubits_Verstraete_2003}. Furthermore, non-maximal entanglement may have stronger robustness to noise, making them optimal \cite{Optimal_Bell_Tests_Do_Not_Require_Maximally_Entangled_State_Ac_n_2005}.

We feature three observers, each performing the same fixed measurement, entangled and characterized by two parameters. The eigenstates of the measurement operator are:
\begin{align}
\label{eq:uw_POVMs}
    u|00\rangle + \sqrt{1-u^{2}}|11\rangle, \quad 
    \sqrt{1-u^{2}}|00\rangle - u|11\rangle,\nonumber\\
    w|01\rangle + \sqrt{1-w^2}|10\rangle, \quad
    \sqrt{1-w^2}|01\rangle - w|10\rangle. 
\end{align}
We observe the maximal genuine network nonlocality for Bell states for measurements characterized by (see Fig.~\ref{fig:box1}) 
\[ (u^2,w^2) = (0.550,0.875) \quad {\rm and} \quad (0.875,0.550).\] 
For these measurements, we have $d\simeq 0.02$ and the corresponding operators are not maximally entangled, and they are different from the standard Bell basis measurements $(u^2,w^2) = (0.500,0.500)$. The optimal measurements we find are also different from the previously known measurements, which, when restricted to a single parameter, $u$, were characterized by the measurements $(u^2,w^2) = (0.63,1)$ or $(u^2,w^2) = (0.83,1)$ given by \cite{Tamas_2020}. In addition, the range of points where $w^2$ and $u^2$ are the same has a measure of zero genuine network nonlocality. 

\begin{result} We consider a more general form of the shared state between the parties and a general form of the measurements that was not considered before, and our study shows that a new set of non-maximal measurements gives the best value of nonlocality $d = 0.019$ compared to the previously known measurements that had $d = 0.008$ \cite{Tamas_2020}. 
\end{result}

It is intriguing how the network topology prefers non-maximal measurement settings in comparison to standard Bell scenarios. Even when the sources distribute pure states, the optimal measurements are those that lie between all Bell-basis projectors, both within the $\{ |00\rangle,|11\rangle\}$ and $\{ |01\rangle,|10\rangle\}$ subspaces, as in Eq.~(\ref{eq:uw_POVMs}) with $(u^2,w^2) = (0.550,0.875)$ \& $(0.875,0.550)$. This can be attributed to the network topology transforming the global quantum state.

Next, we navigate the class of mixed $X$-states using the parameters $p$, $q$, $r$, and $s$, where $r$ is $0.5-p$, giving a total of three effective parameters, where we can visualize a tetrahedron of valid quantum states in the coordinate space (as shown in Fig.~\ref{fig:tetrahedron}). The corners of the tetrahedron are pure states, and the rest are mixed states with the maximally mixed state at the center. We start with the simpler case of mixed $X$-states with a single parameter $p$ (where $q = p$, and $r = s = 0.5-p$); these form two of the edges of the tetrahedron. This scenario can also be visualized by means of a visibility parameter $v$ where the state transitions from one state $\psi_{1}$ to another state $\psi_{2}$ using the expression $v \psi_{1} + (1-v) \psi_{2}$, naturally $p = q = v/2$, where $v \in [0,1]$ and $p,q \in [0,0.5]$.

Interestingly, while considering (see Fig.~\ref{fig:box2}) states with density matrices with a maximum rank of $8$ (rank 2 density matrices for each of the three shared states), we find that the framework automatically learns over these edges without ambiguity, and the mixed states exhibit almost no genuine network nonlocality towards the bulk of the set of states. The only mixed states capable of exhibiting the correlations are highly skewed towards the pure states. This suggests that, unlike standard Bell-inequality violation scenarios, genuine network nonlocality is a much stricter set of correlations.

Further expanding on this, we explored the rest of the tetrahedron through the full set of three-parameter ($p, q,$ and $s$) cases of mixed $X$-states along with the same set of entangled measurements ($u$ and $w$) giving maximal value. This allows the global quantum state density matrix to have a rank of up to $64$, and lets us similarly move smoothly over these sets of states by modulating the state parameters. We confirm the same behavior here, with the lack of genuine network nonlocality in the bulk of the space of mixed states, except for those that are close to the pure states. Finally, we look at the line of mixed states from the maximally mixed state at the center of the tetrahedron to the pure Bell states, or the four corners of the tetrahedron. Again, our findings support the earlier results, confirming that network nonlocality is found only close to the manifold of pure states.

 \subsection{Noise Robustness}

\begin{figure}[!htb]
  \begin{center}
    \includegraphics[scale = 0.7]{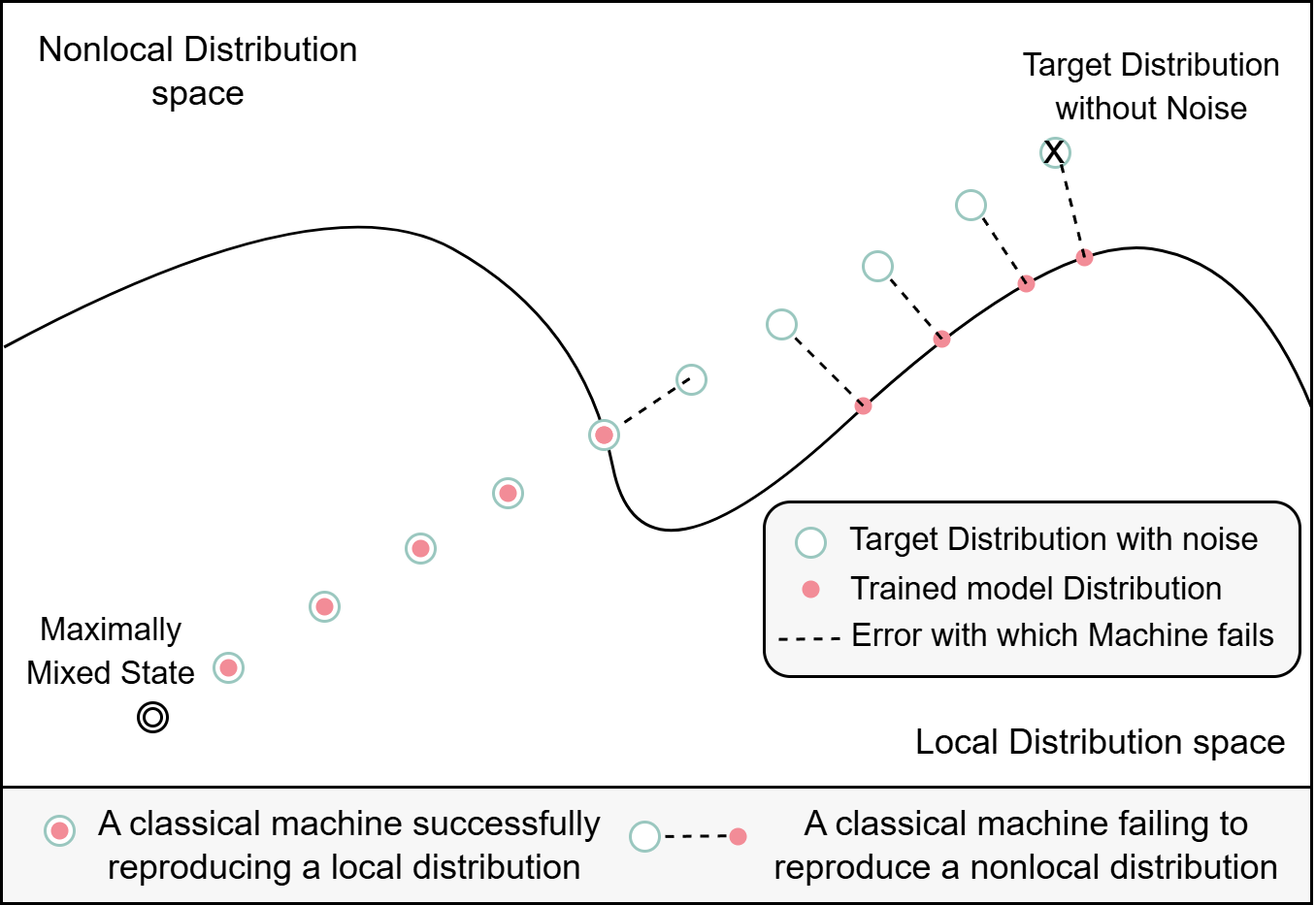}
  \end{center}
  \vspace{-0.25cm}
  \caption{\justifying Adding Werner noise drives any nonlocal distribution towards the local set.}
  \label{fig:Werner_noise_path}
\end{figure}

We explore two noise models to look into the noise robustness of genuine network nonlocality. 
\begin{enumerate}
\item[a)] noise at the sources, and 
\item[b)] noise at the detectors.
\end{enumerate}
This study also benchmarks our model as an effective technique to study quantum network scenarios. We introduce Werner noise with a visibility parameter $v$, such that all three states share the same quantum states and have the form:
\begin{align}
\label{eq:werner state}
    \rho(v) = v|\psi^-\rangle\langle\psi^-| + (1-v)\mathbbm{1}/4,
\end{align}
where $\mathbbm{1}/4$ denotes the maximally mixed state of two qubits.  By adding Werner noise to the quantum state, we analyze the amount of noise it takes for the distribution to enter the local set (see Fig.~\ref{fig:Werner_noise_path}). We vary the visibility parameter for Werner noise for this purpose, transitioning from a maximally mixed state to the chosen quantum state, $|\psi^-\rangle$.

\begin{result}
    We find that the nonlocality measure becomes non-zero with the range $\nu  \in (0.94 , 1]$, for the best state and measurement that we obtained, while the previously known result shows nonlocality from $\nu \geq 0.91$. In other words, we obtain classical models for a range of robustness parameters that were not known before.
\end{result}

\begin{figure}[t]
\centering
\begin{tikzpicture}

\node (a) at (0.5,0)
  {\includegraphics[width=0.85\columnwidth]{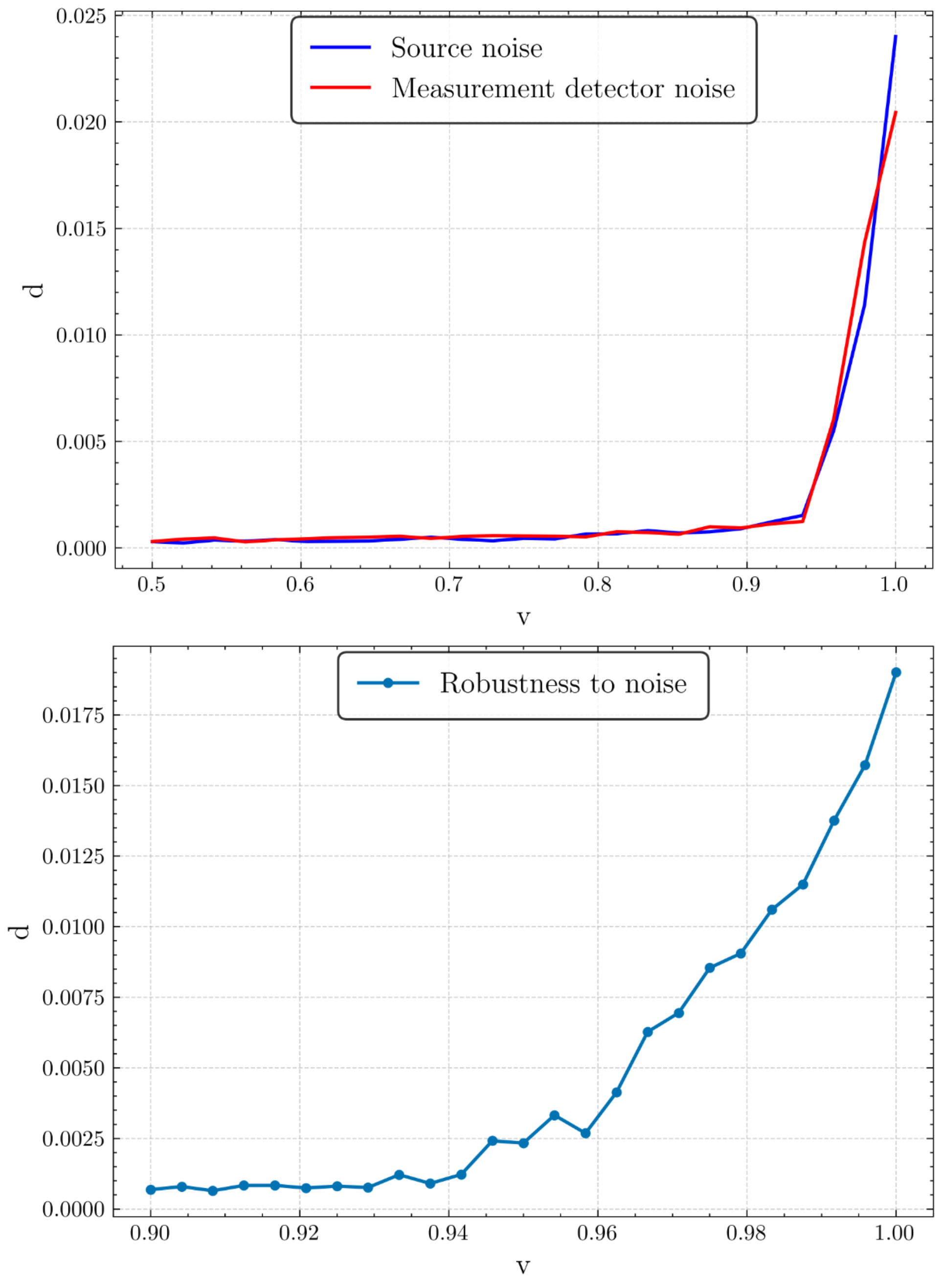}};

\node at (-3,0.2) {\scriptsize{\textbf{(a)}}};
\node at (-3,-4.95) {\scriptsize{\textbf{(b)}}};

%\node at ($(a.north west)+(0.15,-0.15)$) {\textbf{(a)}};
%\node at ($(b.north west)+(0.15,-0.15)$) {\textbf{(b)}};

\end{tikzpicture}
\vspace{-0.25cm}
\caption{\justifying Robustness to Werner noise: (a) We find a noise robustness up to a visibility $v=0.94$ for Werner states using optimal nonlocal measurement settings of $(u^2, w^2) = (0.875,0.550)$. Here $v$ denotes the visibility parameter and $d$ represents the Euclidean distance from the local set. (b) Behavior in the noise threshold regime near $v \in [0.9,1.0]$.}
\label{fig:noise_robustness}
\vspace{-0.25cm}
\end{figure}

The expression of genuine network nonlocality is skewed towards maximally entangled, non-noisy states, with almost no expression in the bulk of the noisy/mixed states. We get a noise robustness of $0.94$, from which we have a gradual increase in the genuine network nonlocality expression, peaking at the maximally entangled Bell state with non-maximally entangled measurement settings with parameters $(u^2,w^2 = 0.875, 0.550)$.  We get the same result with Werner noise in the detector measurement settings, where the POVMs themselves are subjected to noise. The peak in Fig.~\ref{fig:noise_robustness} suggests the genuine network nonlocality expression begins from the $v$ parameter $0.94$. We get this correspondence with both the noise robustness for source and detector settings, as expected.

We further zoomed in with a visibility range $v \in [0.5, 1]$ where we find clear local hidden variable descriptions for the $[0.5,0.938]$ mixed-state distributions, possibly with LHV descriptions existing for $[0.5,0.95)$ (see Fig.~\ref{fig:noise_robustness}). This shows that genuine network nonlocal correlations were much stricter than understood before, and our framework obtains a new clear noise robustness limit.

While earlier work \citep{Renou_Genuine2019:PhysRevLett.123.140401, Tamas_2020, boreiri2024noiserobustproofsquantumnetwork} had commented on the noise robustness of the distributions, we confirm our results on the noise robustness of the triangle scenario using the distribution with $v = 0.938$ visibility, which had previously been presumed to violate any LHV description. In Table I~(\ref{fig:Target_vs_model_distribution}) we give the target RGB4 \cite{Renou_Genuine2019:PhysRevLett.123.140401} distribution with the Bell state $|\phi^+\rangle = \frac{1}{2}(|00\rangle + |11\rangle)$ with noise visibility $v=0.938$ and the entangled basis measurements in~(\ref{eq:uw_POVMs}) followed by the predicted distribution that the model was able to reproduce. In support of our case, the machine was able to approximate the distribution with a discrepancy measure (Euclidean measure) of $d=.0011983$, which is within the error range of the model. In comparison, for the case with zero noise, the value is $d=0.02$. We then recovered the values of the deterministic response functions of Alice, Bob, and Charlie for this case, with their product giving the initial target distribution using the expression~(\ref{eq:network_trilocal_model}). This shows that there exist local realistic descriptions for distributions from a slight Werner noise of $6\%$, and this limit goes far beyond any existing studies on noise robustness. The numerical part can be found in the Code section. 

This behavior is quite interesting, as we again find genuine network nonlocality to be a much stricter set of correlations compared to standard bipartite Bell scenarios. Further, it is really intriguing to find the place of learning algorithms in giving elegant solutions to problems that are difficult to traverse analytically.

\begin{table}[!h]
\centering
\begin{tabular}{l l}
\toprule
 \multicolumn{2}{c}{ Target and model probability distributions with Euclidean} \\ \multicolumn{2}{c}{ distance measure for noise visibility $0.94$ ($10^{-2}$units)} \\
\midrule
Target  &
$\bigl(\!\!$ 0.276,$\!$ 0.276, 4.990,$\!$ 0.767, 0.276, 0.272,$\!$ 1.135, \\ 
Distribution  &  4.508,  4.139, 1.135, 0.327, 0.210, 1.618,  4.508, \\
        &  0.237, 0.327, 0.272, 0.276, 1.135, 4.508,  0.276, \\
        & 0.276, 4.139, 1.618, 1.135, 4.990, 0.327,  0.237, \\
        & 4.508, 0.767, 0.210, 0.327, 4.508, 0.767,  0.327, \\
        &  0.210, 1.618, 4.508, 0.327, 0.237,  0.327,  0.327,\\
        & 8.384, 0.508, 0.237, 0.210, 0.508, 1.999, 1.135, \\
        &  4.990, 0.237, 0.327, 4.139, 1.135, 0.210, 0.327, \\
        &   0.210, 0.237, 0.508, 1.999, 0.327, 0.327,  1.999, \\
        & 6.894 $\!\!\bigr)$ \\
\midrule
Model &
$\bigl(\!\!$ 0.272, 0.276,$\!$ 4.976, 0.775, 0.255,$\!$ 0.254,$\!$ 1.147, \\
Distribution & 4.485,  4.102, 1.201, 0.329, 0.218, 1.621, 4.454, \\
        & 0.248, 0.332, 0.266, 0.268, 1.140, 4.466, 0.275, \\ 
        &  0.282, 4.091, 1.660, 1.150, 4.988, 0.320, 0.246, \\
        & 4.483, 0.784, 0.202, 0.322, 4.494, 0.767, 0.317, \\
        &  0.206, 1.628, 4.467, 0.305, 0.238, 0.330, 0.316, \\
        & 8.374, 0.505, 0.239, 0.222, 0.596, 2.016, 1.186, \\
        &  4.946, 0.232, 0.329, 4.190, 1.148, 0.229, 0.326, \\
        & 0.208, 0.234, 0.479, 1.985, 0.328, 0.325, 2.044, \\
        & 6.899 $\!\!\bigr)$ \\
\midrule
 \multicolumn{2}{c}{ $\|p_{\text{target}} - p_{\text{model}}\|_2 = 
0.00198311$} \\
\bottomrule
\end{tabular}
\caption{\justifying $64$ element sequential array of target and model probability distribution for noise visibility $0.94$}
\label{fig:Target_vs_model_distribution}
\vspace{-0.5cm}
\end{table}

\subsection{Genuine Network Nonlocality with Dissimilar Entanglement and Noise}

To further understand these correlations, we explore the case of dissimilar sources for the triangle network scenario. Specifically, by fixing a few sources and varying the others from maximally mixed states to maximally entangled ones.

We explore three classes of quantum states with different combinations of dissimilar states, given by:
\begin{align} \nonumber
    %\rho^{MMX} &= \rho^{M}\otimes\rho^{M}\otimes\rho^{X} \\ \nonumber
    \rho^{{\rm MEX}} &= \rho^{M}\otimes\rho^{E}\otimes\rho^{X}, \\ \nonumber
    \rho^{{\rm EXX}} &= \rho^{E}\otimes\rho^{X}\otimes\rho^{X}, \\ \nonumber
    \rho^{{\rm EEX}} &= \rho^{E}\otimes\rho^{E}\otimes\rho^{X}, 
    \label{eq:dissimiliar_states}
\end{align}
where $\rho^X$ is the $X$-state density matrix given in Eq.~\eqref{eq:pqs_matrix}, which varies from $\rho^M$ to $\rho^C$ to $\rho^E$, with
\[ \rho^{M} \!\! = \!\!
\frac{1}{4}\!\!\begin{bmatrix}
1 & 0 & 0 & 0\\
0 & 1 & 0 & 0\\
0 & 0 & 1 & 0\\
0 & 0 & 0 & 1
\end{bmatrix}\!\!, \, 
\rho^{E} \!\! =
\frac{1}{2} \!\!\begin{bmatrix}
1 & 0 & 0 & 1\\
0 & 0 & 0 & 0\\
0 & 0 & 0 & 0\\
1 & 0 & 0 & 1
\end{bmatrix}\!\!, \, 
\rho^{C} \!\! = 
\frac{1}{2} \!\!
\begin{bmatrix}
1 & 0 & 0 & 0\\
0 & 0 & 0 & 0\\
0 & 0 & 0 & 0\\
0 & 0 & 0 & 1
\end{bmatrix}\!\!.\]

\begin{result}
We find that genuine network nonlocal correlations require a certain degree of entanglement in all sources to necessitate network nonlocality. In addition, we present a separate class of non-maximally entangled states distinct from Werner states capable of exhibiting these correlations.
\end{result}

\begin{figure}[!t]
     \centering
     
    \begin{tikzpicture}
    \node (a) at (0.5,0)
      {\includegraphics[width=0.85\columnwidth]{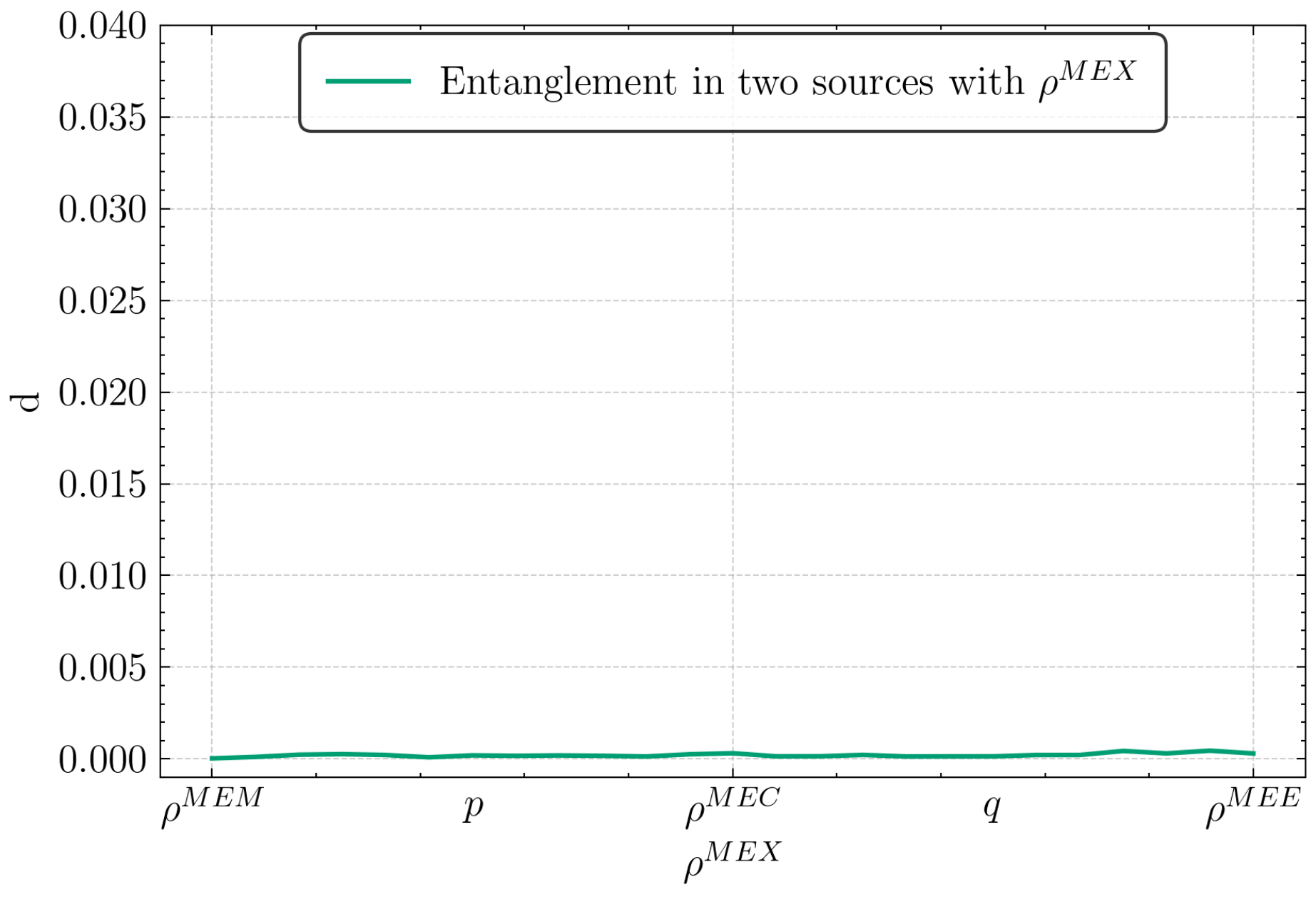}};
    
    \node (b) at (0.5,-0.6\columnwidth)
      {\includegraphics[width=0.85\columnwidth]{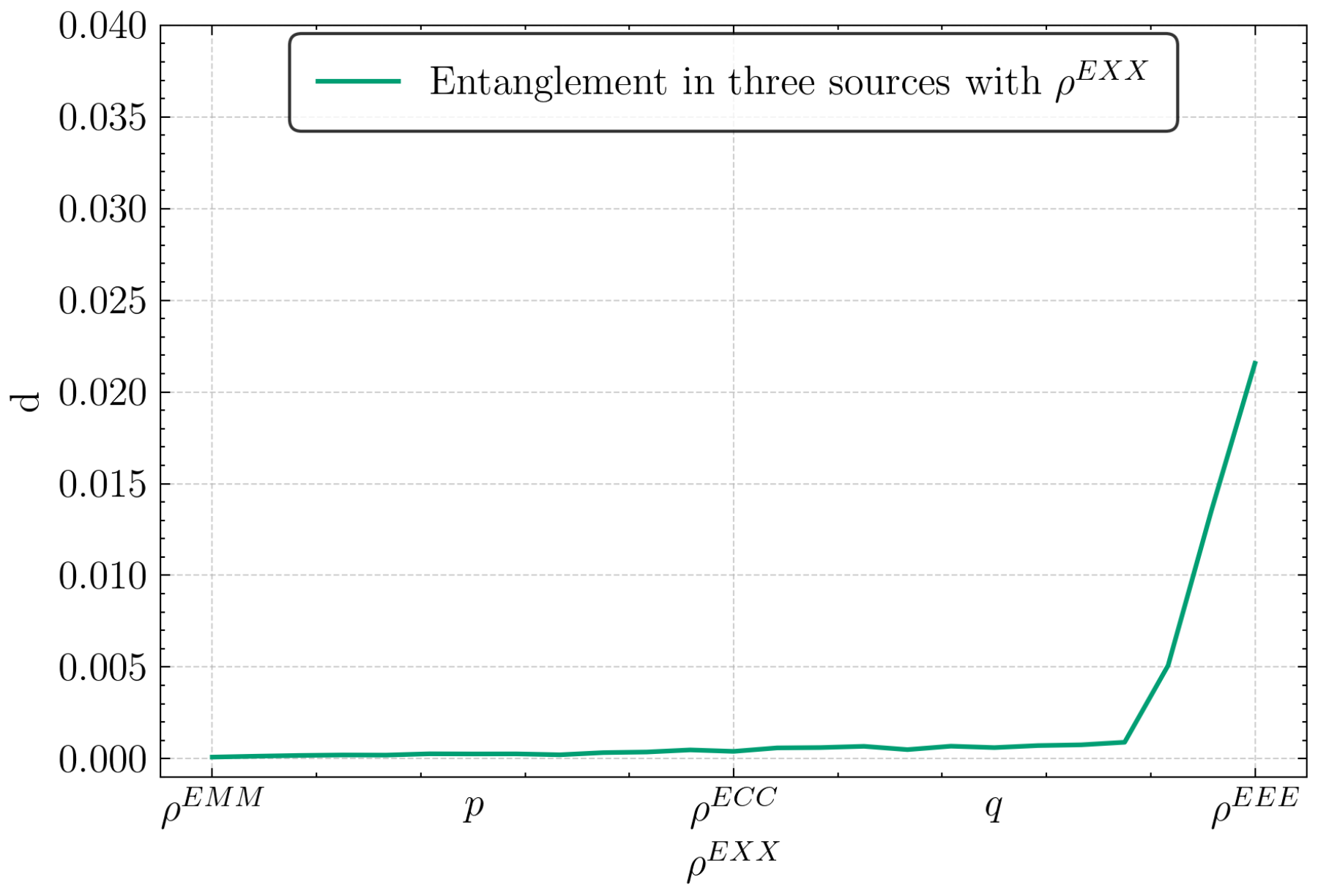}};

    \node (c) at (0.5,-1.2\columnwidth)
      {\includegraphics[width=0.85\columnwidth]{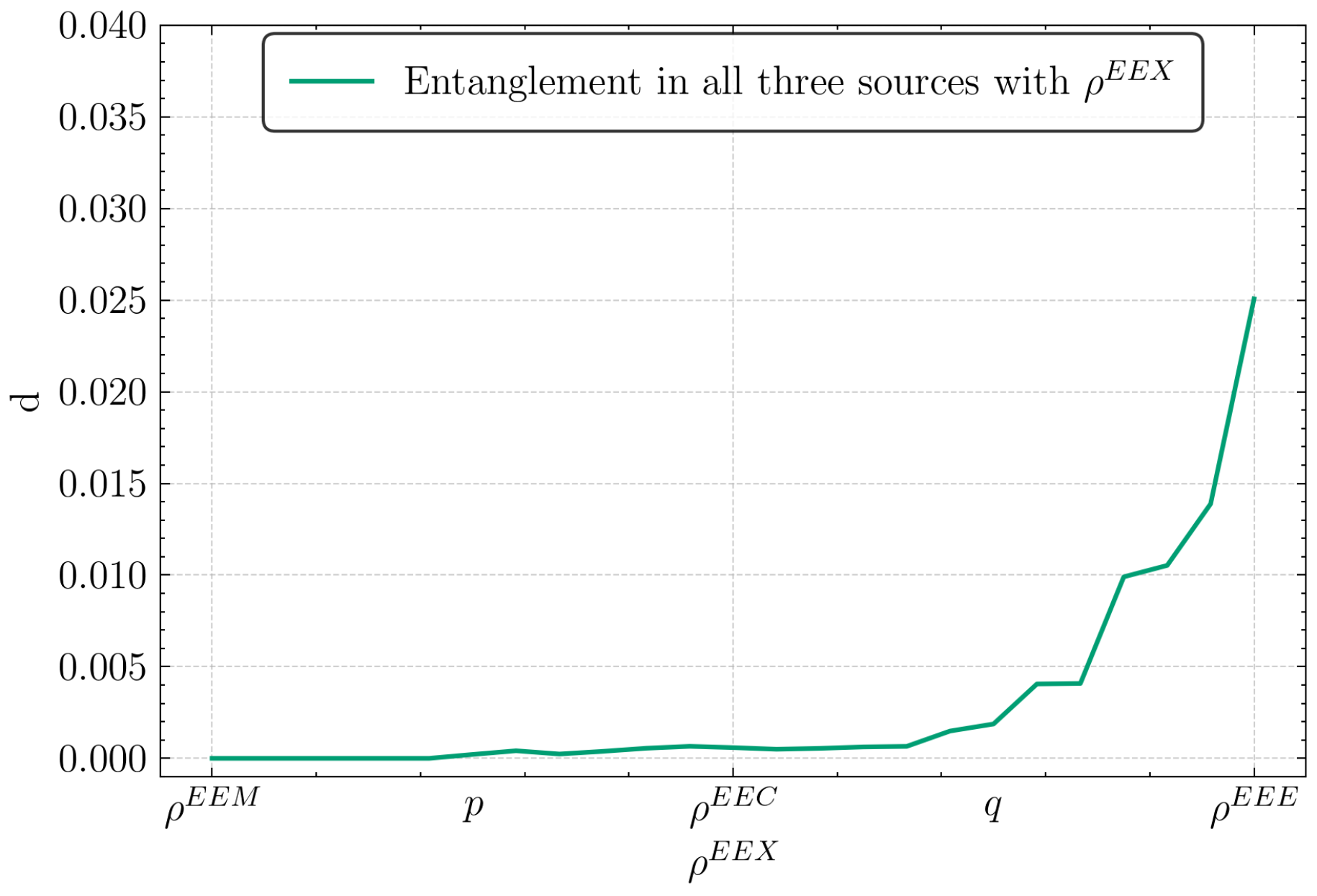}};
    
    \node at (-3.2,-2.3) {\scriptsize{\textbf{(a)}}};
    \node at (-3.2,-7.5) {\scriptsize{\textbf{(b)}}};
    \node at (-3.2,-12.6) {\scriptsize{\textbf{(c)}}};

    \end{tikzpicture}
    \vspace{-0.25cm}
    \caption{\justifying Genuine network nonlocality (GNN) under variation of $\rho^X$ in (a) $\rho^M\otimes\rho^E\otimes\rho^X$, (b) $\rho^E\otimes\rho^X\otimes\rho^X$, (c) $\rho^E\otimes\rho^E\otimes\rho^X$. The state $\rho^X$ is varied from $\rho^M$ to $\rho^C$ via the parameter $p$ and from $\rho^C$ to $\rho^E$ using the parameter $q$ in~(\ref{eq:pqs_matrix}). The vertical axis $d$ denotes the Euclidean distance of the target distribution from the local set. The optimal measurement settings occur at parameters $(u^2,w^2) = (0.875,0.550)$.}
    \label{fig:dissimilar_sources}
    \vspace{-0.25cm}
\end{figure}

The framework successfully identified local models for all states of the class $\rho^{{\rm MEX}}$ (see Fig.~\ref{fig:dissimilar_sources}a), suggesting a lack of genuine network nonlocality with only one ($\rho^{{\rm MMX}}$) or two entangled states ($\rho^{{\rm MEX}}$). However, for $\rho^{{\rm EEX}}$ where two sources are entangled and the third transitions from separable to entangled, a local model was found only when the third source remained below a certain degree of entanglement (see Fig.~\ref{fig:dissimilar_sources}c), i.e., once $X$ began incorporating entangled states beyond a certain degree, no LHV descriptions were found. We further strengthened this result by studying the class of mixed states from $\rho^{{\rm EXX}}$ (see Fig.~\ref{fig:dissimilar_sources}b). This suggests a sufficient degree of non-zero entanglement is required in all sources to facilitate these correlations.

This further supports the results of \cite{sekatski2022partialselftestingrandomnesscertification} on all sources requiring a minimum entanglement. So for genuine network nonlocality in the triangle network system, all its states need to have a certain degree of entanglement with entangled measurements, where the maximally entangled states, coupled with specific entangled measurements, $(u^2,w^2) = (0.875,0.550)$ give maximal nonlocality.

Aside from being sufficiently entangled, these states that lead to genuine network nonlocality, from the class of $\rho^{{\rm EEX}}$ and $\rho^{{\rm EXX}}$ mixed states where $\rho^{X}$ transitions from the classically correlated state $\rho^{C}$ to the maximally entangled one $\rho^{E}$, are non-maximally entangled states distinct from Werner states. We see this behavior with the class of $\rho^{XXX}$ mixed states with similar sources (Fig.~\ref{fig:tetrahedron}), but here with dissimilar sources we increased their robustness (see Fig.~\ref{fig:dissimilar_sources}) by fixing maximally entangled sources for one ($\rho^{{\rm EEX}}$) or two ($\rho^{{\rm EXX}}$) sources in the network.

\begin{figure}[!ht]
     \centering
     
    \begin{tikzpicture}
    \node (a) at (0.5,0)
      {\includegraphics[width=0.85\columnwidth]{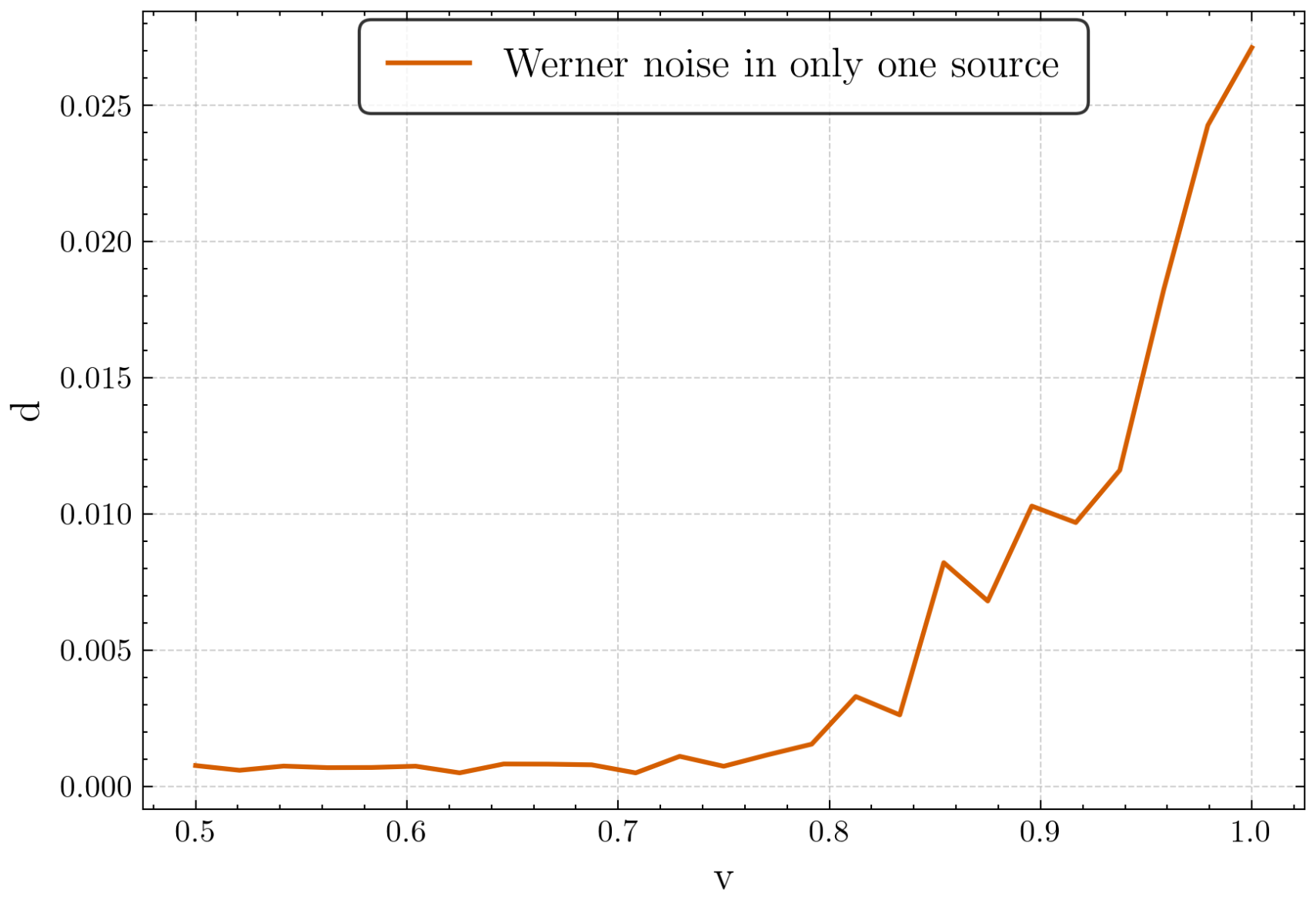}};
    
    \node (b) at (0.5,-0.6\columnwidth)
      {\includegraphics[width=0.85\columnwidth]{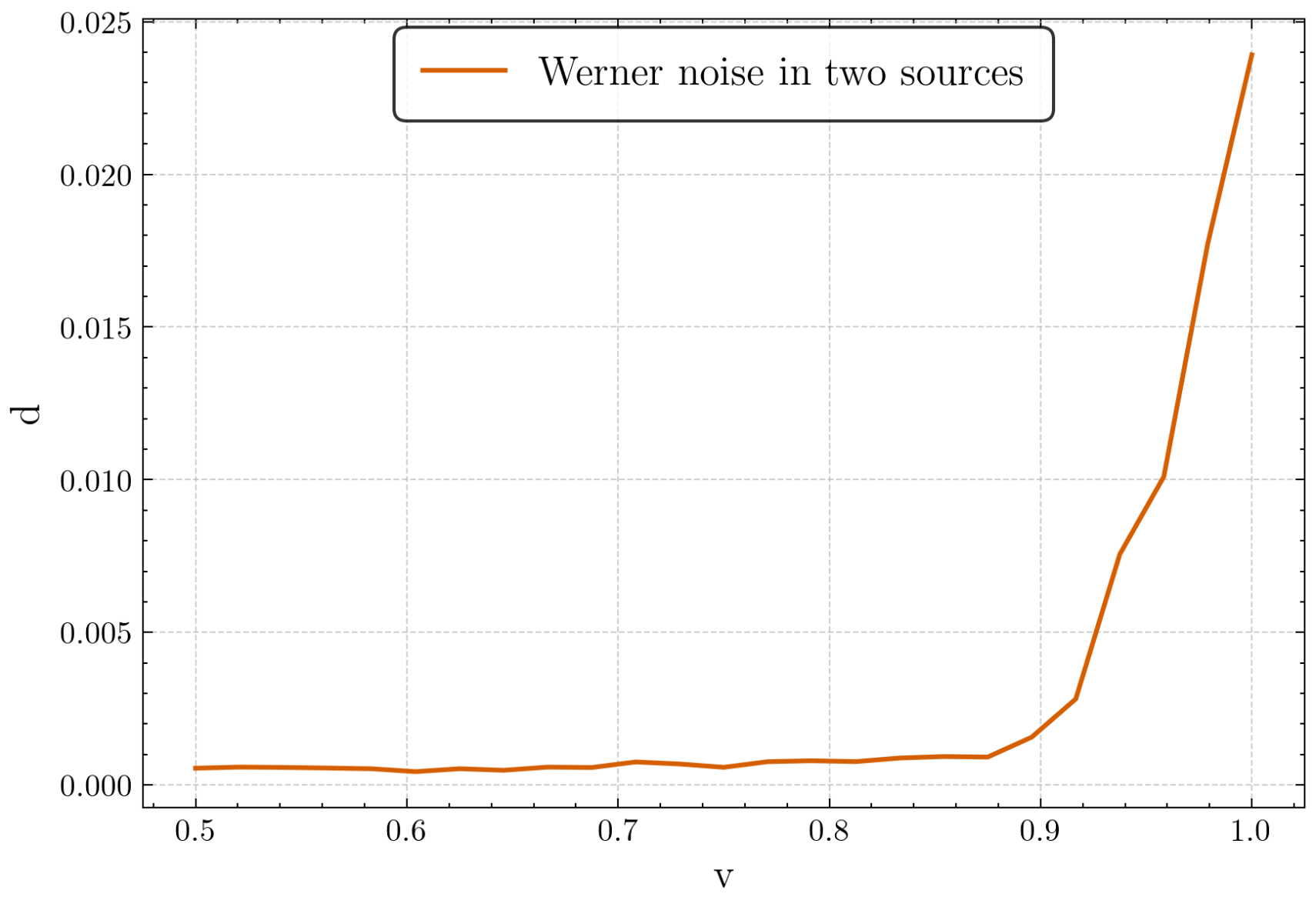}};

    \node (c) at (0.5,-1.2\columnwidth)
      {\includegraphics[width=0.85\columnwidth]{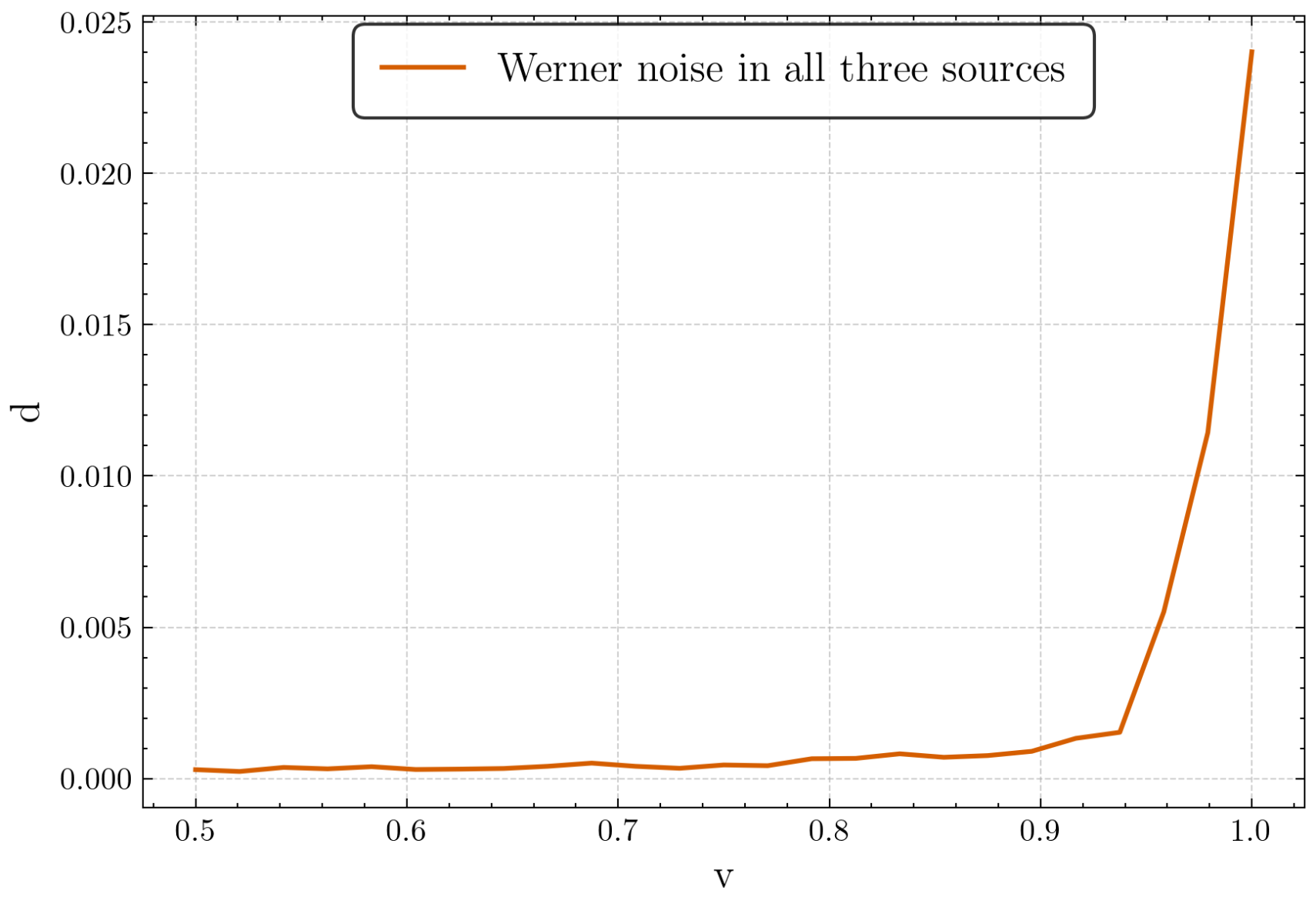}};
    
    \node at (-3.2,-2.3) {\scriptsize{\textbf{(a)}}};
    \node at (-3.2,-7.5) {\scriptsize{\textbf{(b)}}};
    \node at (-3.2,-12.6) {\scriptsize{\textbf{(c)}}};

    \end{tikzpicture}
    \vspace{-0.25cm}
    \caption{\justifying Werner noise is added to (a) a single source, (b) two sources, and (c) all three sources. The optimal measurement settings are $(u^2,w^2) = (0.875,0.550)$. The horizontal axis $v$ denotes the Werner visibility parameter, and $d$ represents the Euclidean distance from the local set.}
    \label{fig:dissimilar_noisy_sources}
    \vspace{-0.25cm}
\end{figure}

Studies on non-symmetric scenarios can also clarify both the viability and limitations of the current model. The intermediate regions between the family of distributions with all sources entangled and those with all sources separable extend well beyond the descriptive capacity of the fully separable model that underlies our layered LHV neural network architecture~(\ref{eq:mixed_separable_distribution}). At present, we lack a general separable decomposition framework for quantum networks. This implies that in scenarios involving three sources, one or two sources may be entangled while the overall distribution could still admit an LHV description. But quite interestingly, our current layered LHV framework captures an elegant expression transitioning over these families of distributions. This is especially intriguing because a fully separable model lacks the degree of freedom to capture these subtler interplays. This suggests nuances in our understanding of genuine network nonlocality and the separable decompositions of network scenarios, and this could very well lead to interesting descriptions for these correlations.

Further delving into the case of noise robustness, we tested for genuine network nonlocality with dissimilar noisy sources; we considered three scenarios: first, by adding Werner noise in one state; next, with two states; and finally, with all three states, which is the same as the standard noise robustness study in the previous session. We find that the quantum network with a single noisy source in Fig.~\ref{fig:dissimilar_noisy_sources}a has much better noise robustness as expected, followed by the case with two noisy sources in Fig.~\ref{fig:dissimilar_noisy_sources}b, and finally the least when all three are noisy in Fig.~\ref{fig:dissimilar_noisy_sources}c with a noise robustness limit $v = 0.94$ beyond which genuine network nonlocality starts to arise.

With these results, we can conjecture the genuine network non-local correlations in the triangle scenario to be close to the local set, since with the slightest addition of noise, the distribution falls into the local set. Referencing the earlier figure in Fig.\ref{fig:Local_and_nonlocal_domain}, we should find these correlations close to the boundary of the local set of quantum network scenarios with independent sources. And, distributions from network scenarios with Bell states having maximal entanglement, coupled with the optimal set of entangled basis measurements $(u^2,w^2) = (0.875,0.550)$~(\ref{eq:uw_POVMs}) can be found at the maximum distance from the local set.

\begin{figure*}[ht]
\centering
\begin{tikzpicture}

\node (a) at (0,0)
  {\includegraphics[width=0.85\columnwidth]{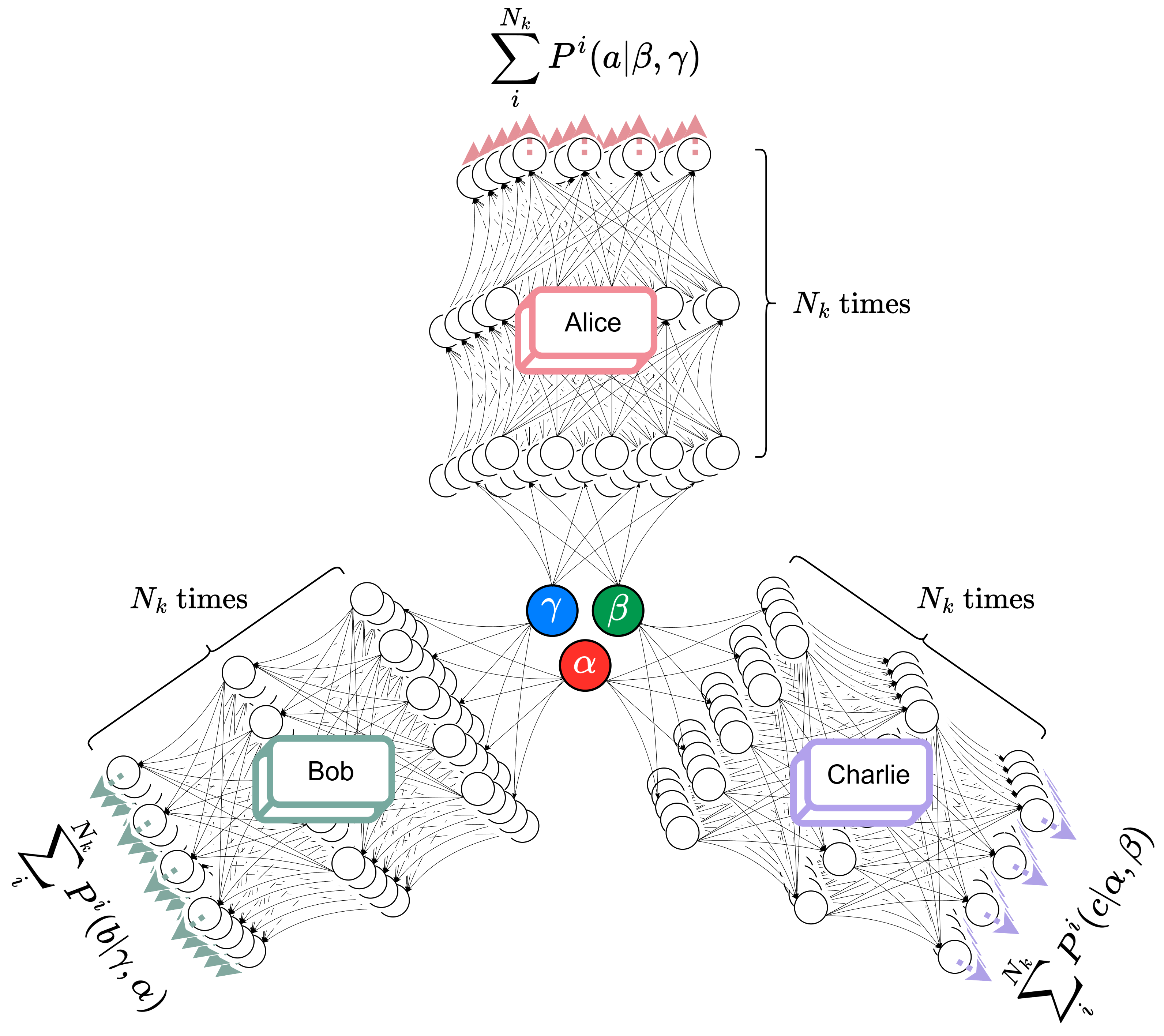}};

\node (b) at (\columnwidth,0)
  {\includegraphics[width=1.15\columnwidth]{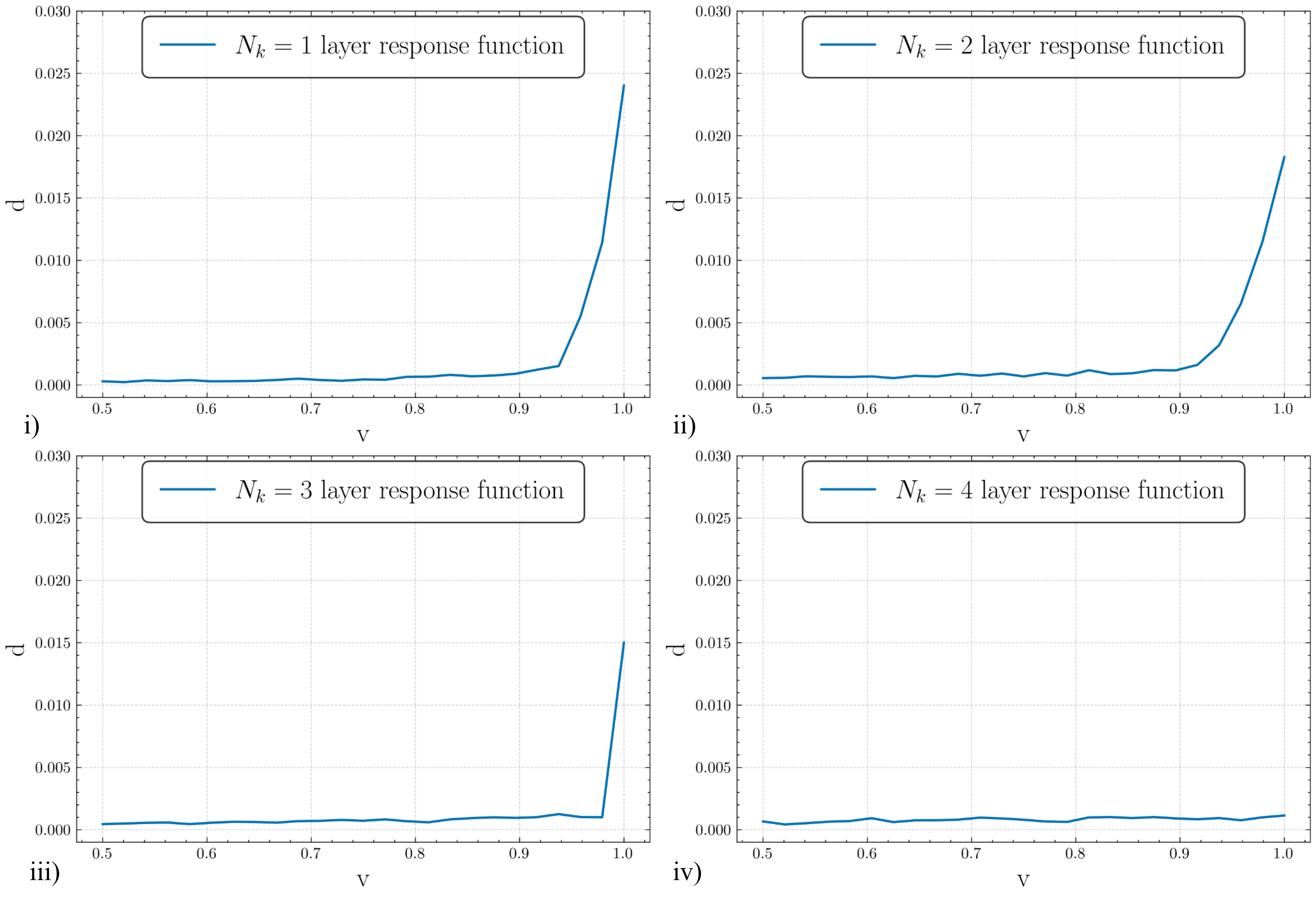}};

\node at (0,-3.8) {\scriptsize{\textbf{(a)}}};
\node at (9.0,-3.8) {\scriptsize{\textbf{(b)}}};

%\node at ($(a.north west)+(0.15,-0.15)$) {\textbf{(a)}};
%\node at ($(b.north west)+(0.15,-0.15)$) {\textbf{(b)}};

\end{tikzpicture}
\vspace{-0.35cm}
\caption{\justifying Layered LHV neural network with shared randomness: (a) Architecture consisting of $N_k$ neural network layers trained in parallel, incorporating classical shared randomness to learn the target distribution. (b) No local hidden variable (LHV) model is found for distributions with shared randomness sizes $N_k \in \{1,2,3\}$, whereas LHV descriptions exist for $N_k \ge 4$. Further, the network nonlocal correlations persist in noisy scenarios as well. We use the optimal measurement settings of $(u^2,w^2) = (0.875,0.550)$. Here, $v$ denotes the Werner visibility parameter and $d$ the Euclidean distance from the local set.}
\label{fig:shared_randomness}
\vspace{-0.25cm}
\end{figure*}

\section{Robustness of genuineness to shared randomness}

These network correlations are distinguished by their genuineness, which sets them apart from standard Bell nonlocal correlations. In this setting, nonlocality arises jointly from all sources, making genuine network nonlocality a stricter notion than genuine multipartite nonlocality.

While the Bell test relies on the fundamental assumptions of realism, locality, and measurement independence (or “free will”), the network scenario instead replaces the freedom-of-choice assumption with a source independence assumption. Within this framework, we obtain nonlocal correlations that form a significantly stricter set than those arising in standard Bell scenarios.

We relax the assumption of source independence in the local model. Testing this assumption is crucial, since in practice it is rarely possible to guarantee that experimental sources are fully independent. We therefore study whether there is robustness to dependence of the sources by introducing a shared randomness into the network scenario. It essentially quantifies how adversarial-proof the network’s quantum correlations are. A network that maintains nonlocal correlations even when shared randomness is introduced remains verifiably quantum under weaker trust assumptions. 

Further, we want to answer whether this robustness is restricted to ideal pure states or whether it also persists for slightly noisy mixed states as well, which are more relevant to realistic experimental conditions.

The probability distribution in the network scenario with fixed inputs and outputs $a, b,$ and $c$, constrained by the source independence assumption on the hidden variables $\lambda_1, \lambda_2$, and $\lambda_3$, while admitting a classical shared randomness, can be written as:
\begin{eqnarray}
\label{eq:source_independence}
    p'(a,b,c)  &= &   \sum_{k=1}^{N_k}  p(k)  \int d\lambda_1 d\lambda_2 d\lambda_3  \nonumber \\
    && \qquad \times \, \rho_1(\lambda_1,k) \rho_2(\lambda_2,k) \rho_3(\lambda_3,k) \nonumber \\ 
    && \qquad  \times \,P(a|\lambda_2, \lambda_3,k) P(b|\lambda_1, \lambda_3,k) \nonumber \\
    && \qquad \qquad \times \, P(c|\lambda_1, \lambda_2,k), 
\end{eqnarray}
where the shared variable $k$ takes $N_k$ values. Here, $N_k$ is the possible cases of classical shared randomness, and $k$ is the variable.

Using the neural network framework for the layered response function, we introduce a dependence to the response of three observers, Alice, Bob, and Charlie, which does not go as far as classical communication (signaling) but introduces a shared randomness that the observers can use (see Fig.~\ref{fig:shared_randomness}). As seen in~(\ref{eq:source_independence}), there is the classical randomness that Alice, Bob, and Charlie share; now, depending on the value of $N_k$, we can introduce controlled shared randomness $k$ into the network scenario. We achieve this by adapting our earlier expression~(\ref{eq:mixed_separable_distribution}) for the local-realistic hidden variable description in network scenarios $p(a,b,c)$ to include a shared variable/degree of freedom $k$ in the individual response function, which can be represented as:
\begin{align}\label{eq:shared_randomness}
p'(a,b,c) &= \sum_{i,j,l=1,1,1}^{k_{a} k_{b} k_{c}} \lambda^{\alpha}_i\lambda^{\beta}_j\lambda^{\gamma}_l \ {\rm Tr}(P_{A_5A_4}^a\rho_{A_5A_4}^{i,j,l}) \nonumber \\
& \qquad  \times \, {\rm Tr}(P_{B_1B_6}^b\rho_{B_1B_6}^{i,j,l}) {\rm Tr}(P_{C_3C_2}^c\rho_{C_3C_2}^{i,j,l}) \nonumber \\
&= \sum_{k=1}^{N_k} \lambda^{\alpha}_i\lambda^{\beta}_j\lambda^{\gamma}_l {\rm Tr}(P_{A_5A_4}^a\rho_{A_5A_4}^{k}) \nonumber \\
& \qquad  \times \, {\rm Tr}(P_{B_1B_6}^b\rho_{B_1B_6}^{k}) {\rm Tr}(P_{C_3C_2}^c\rho_{C_3C_2}^{k}),
\end{align}    
here, the combined coefficients $\lambda^{\alpha}_i\lambda^{\beta}_j\lambda^{\gamma}_l$ are absorbed into the neural network parameters and implicitly accounted for during training.

Now the question we ask is if there exist local realistic strategies that the observers can apply with the classical shared randomness resource to compromise genuine network nonlocal correlations. The possibility is particularly intriguing from the perspective of device-independent security: it suggests that genuine network nonlocality may persist even in the presence of adversarial entities capable of compromising security using only local operations and shared randomness (LOSR).

\begin{result}
We find that genuine network nonlocal correlations exhibit some robustness to shared randomness even with noise. In particular, there exists no local realistic description of the quantum correlations with just 2 or 3 bits of classical shared randomness between the sources. 
\end{result}

Our analysis shows that for $N_k=2$ and $N_k=3$ layer network configurations, genuine nonlocality persists; however, with the addition of a fourth layer in $N_k=4$, the shared randomness resource becomes sufficient for a local hidden variable description to fully reproduce the correlations, causing the genuineness to disappear. Moreover, by introducing shared randomness to our noisy distributions under Werner noise, we find that this robustness is not restricted to pure states. Slightly noisy mixed states are also able to sustain genuine nonlocal correlations, which is encouraging from the perspective of practical implementations.

We also explored the case of partial shared randomness, where only two of the sources share randomness while the third remains independent. In this setting, we again observe that some robustness to genuineness is retained, reinforcing the idea that network-based nonlocal correlations can tolerate imperfections in source independence to a nontrivial extent.

\section{Discussion}

The interdisciplinary field of quantum resources and quantum frameworks with algorithms in machine learning presents an emerging and highly fascinating direction of research. In recent years, machine learning has shown considerable promise in interdisciplinary quantum foundations research. It has become part of an ever-expanding toolbox of techniques in quantum information theory, including applications such as the formulation of Bell-type inequalities and the characterization of quantum correlations \cite{Machine_Learning_Nonlocal_Correlations:PhysRevLett.122.200401, Machine_Learning_Detection_of_Bell_nonlocality_in_Quantum_Many-Body-Systems:PhysRevLett.120.240402, Tamas_2020, Probabilistic_Graphical_Models:Koller1989, Learning_Functional_Causal_Models_with_Generative_Neural_Networks:Goudet2018, Causality:Hitchcock2001}. However, going deeper beyond the common practice of employing machine learning purely as a computational tool, we find a developing yet important niche direction where machine learning architectures can be adapted using physics and foundational insights to build novel and more robust frameworks.

On the same note, machine learning algorithms are traditionally known to be useful for learning and predicting from structured data and hence are seldom used in providing conclusive answers to foundational questions. The framework we use is different and lies in the interdisciplinary niche, where, though algorithmic in nature, our framework is structurally foundational, as it succeeds by answering the question of whether the Bell experiment statistics can be learned or not. Specifically, our framework employs a quantum-informed LHV Bayesian network, where node values are probability distributions consistent with the classical DAG structure of the quantum network we are studying. Coupled with a learning algorithm, this framework adjusts the node values by learning the local statistics of the network scenario. This fundamental difference raises the utility of the model from a data-driven model to a foundational framework.

Using the framework, we overcome the limitations of existing methods in network scenarios and provide a causally constrained layered LHV learning algorithm to study and characterize ideal and non-ideal scenarios of genuine network nonlocality. Equipped with the $2 \times 2$ layered response function, which has the degrees of freedom to sufficiently capture all possible local hidden variable descriptions of mixed-state distributions, the model remains operational to use.

Next, we considered a general form of the shared state between the parties traversing through both pure states and mixed states and a general form of the entangled basis measurements that was not considered before. Our study revealed a new set of optimal measurement settings that gave the maximum genuine network nonlocality compared to previous measurements \cite{Tamas_2020}. It was further interesting to see that when measuring Bell states, the optimal measurements were not projective measurements in the Bell basis but non-maximal measurements that interpolate between Bell-basis projectors, both within the $\{ |00\rangle,|11\rangle \}$ and $\{ |01\rangle,|10\rangle \}$ subspaces. In other words, when taking a Bell state ($(|00\rangle + |11\rangle)/\sqrt{2}$) as the shared state, optimality requires entangled measurement elements both in the support of the Bell state $\{ |00\rangle,|11\rangle \}$ and in the complementary subspace $\{ |01\rangle,|10\rangle \}$, even though the latter carries no population in the measured state. This can be attributed to the quantum network topology transforming the global quantum state.

Coming to the non-ideal scenarios with the class of mixed $X$-states, we find the GNN expression in the triangle scenario to be skewed towards pure states, with the optimal measurement settings we found giving the best expressivity. Here, we find that the bulk of the mixed-state distribution lacks genuine network nonlocality. Backing this up, we find a new noise robustness value with visibility $0.94$ with Werner noise, below which the distribution always gives a local realistic description; this is also the current best estimate of noise robustness in the triangle scenario. Further, by studying dissimilar sources, we find that genuine network nonlocal correlations require a certain degree of entanglement in all sources to necessitate network nonlocality; this aligns with the observations in \cite{sekatski2022partialselftestingrandomnesscertification}. In addition, we also present a separate class of non-maximally entangled states distinct from Werner states capable of exhibiting
these correlations.

Finally, by adapting the framework, we introduced classical shared randomness into our model to study the robustness of these genuine network nonlocal correlations to shared randomness under local operations and shared randomness (LOSR). And intriguingly, we find that genuine network nonlocal correlations exhibit some robustness to shared randomness even with noise. The robustness essentially quantifies how adversarial-proof the network’s quantum correlations are. A network that maintains nonlocal correlations even when shared randomness is introduced remains verifiably quantum under weaker trust assumptions. As an adversary, distributing these hidden shared variables among sources or parties would still fail to reproduce the correlations predicted by the quantum model.

Looking ahead, an interesting open question concerns why our current layered LHV framework captures an elegant expression transitioning over these families of distributions, despite being based on a fully separable model, which lacks the degree of freedom to capture these subtler interplays. At present, we lack a general separable decomposition framework for quantum networks, and the fact that we got a result suggests that the true separable decomposition might not be far off. Experimenting with these networks to find where the models fail for distributions having a local description might lead us to a solution. This is because fixing this would require the model to have a new DAG, and the mathematical analog of that would give us a general separable decomposition framework. Hence, these could very well lead to interesting descriptions for this novel set of correlations. As an additional tool, LOSR can be used in this study by adapting the model to introduce shared randomness. As genuine network nonlocal correlations are closed under LOSR, we can verify whether the correlations are nonlocal only due to the limitations of the model by checking whether they persist even after introducing shared randomness.

Another intriguing research direction lies in a robust machine learning framework that can give a perfect certificate of nonlocality in Bell scenarios. While semidefinite programming techniques based on the Navascués-Pironio-Acín (NPA) hierarchy offer mathematically sound certificates, machine learning methods are well-suited to identifying structural patterns and features in correlations that are difficult to capture analytically. In a follow-up work, we aim to address this and adapt our framework to incorporate SDP-based techniques for Bell tests. This hybrid strategy could address a central limitation of existing machine learning approaches, namely, the absence of rigorous certificates, while leveraging the most robust numerical machinery currently available for certifying nonlocality. It would be rather interesting if in this way one could derive Bell-type inequalities for scenarios that aren't easily analytically tractable.

Extending the study, establishing an analytical proof of the optimality of these measurement settings remains an important direction for future work. The measurement settings were particularly intriguing owing to their non-maximal nature and the presence of entangled measurement elements both in the support of the state under measurement and in their complementary subspace. Further, figuring out communication complexity protocols for applications with provable quantum speedup from the network scenario's entanglement-based protocols would be a really interesting direction of research.

Machine learning offers valuable insight into the structure and boundaries of correlations, a problem that has long challenged analytical and numerical approaches. In this work, we elevated machine learning from being a computational tool to serving as a foundational framework for studying quantum network scenarios. This shift opens up new layers of complexity, revealing how quantum correlations can be optimized and potentially harnessed in practical applications. At the same time, it highlights how new algorithmic frameworks can be constructed by drawing directly from quantum resources and foundational principles. These developments bridge the gap between numerical learning approaches and rigorous foundational characterizations of nonlocality. Developing and adapting such interdisciplinary frameworks offers promising avenues for achieving a deeper understanding of quantum phenomena and for building more effective methods to simulate and capture the behavior of a quantum nature.

%\vspace{0.5cm}

\section{Data Availability}
The authors declare that the data supporting the findings of this study are available in the article.

%\vspace{-0.15cm}

\section{Code Availability}

Our implementation of the layered LHV neural network framework for the triangle network scenario can be found here: \url{https://github.com/ananthrishna/GNN-LHV-k-triangle.git}. Numerical proof supporting the LHV description for the Werner noise robustness limit $v=0.94$ is also provided.

\begin{acknowledgments}
The authors thank the Quantum Information Theory group and the Quantum Foundations group for fruitful discussions. We acknowledge funding support for the Chanakya-PG fellowship from the National Mission on Interdisciplinary Cyber Physical Systems, of the Department of Science and Technology, Govt. of India, through the I-HUB Quantum Technology Foundation. The authors thank the Padmanabha computational cluster, which was made available through the Center for High-Performance Computation at IISER-Thiruvananthapuram. AS and DS acknowledge support from $\text{STARS (STARS/STARS-2/2023-0809)}$, Govt. of India. AS also acknowledges the support of the National Quantum Mission of the Department of Science and Technology, Government of India, through the Foundation for QC Innovation. 
\end{acknowledgments}

\nocite{*}
\bibliography{mybib}

\end{document}